**Digital exclusion among middle-aged and older adults in China: age–period–cohort**

**evidence from three national surveys, 2011–2022**

Yufei Zhang[1], Zhihao Ma[1]

[1] Computational Communication Collaboratory, School of Journalism and Communication, Nanjing University, Nanjing 210023, China

**Corresponding author:**

Zhihao Ma

Computational Communication Collaboratory

School of Journalism and Communication

Nanjing University

Room 345, Zijin Building, Nanjing University (Xianlin Campus), 163 Xianlin Road, Qixia District, Jiangsu Nanjing, 210023 China

Phone: 86 17561538460

Email: redclass@163.com

# Abstract

Amid China's ageing and digital shift, digital exclusion among older adults poses an urgent challenge. To unpack this phenomenon, this study disentangles age, period, and cohort effects on digital exclusion among middle-aged and older Chinese adults. Using three nationally representative surveys (CHARLS 2011–2020, CFPS 2010–2022, and CGSS 2010–2021), we fitted hierarchical age–period–cohort (HAPC) models weighted by cross-sectional survey weights and stabilized inverse probability weights for item response. We further assessed heterogeneity by urban–rural residence, region, multimorbidity, and cognitive risk, and evaluated robustness with APC bounding analyses. Across datasets, digital exclusion increased with age and displayed mild non-linearity, with a small midlife easing followed by a sharper rise at older ages. Period effects declined over the 2010s and early 2020s, although the pace of improvement differed across survey windows. Cohort deviations were present but less consistent than age and period patterns, with an additional excess risk concentrated among cohorts born in the 1950s. Rural and western residents, as well as adults with multimorbidity or cognitive risk, remained consistently more excluded. Over the study period, the urban–rural divide showed no evidence of narrowing, whereas the cognitive-risk gap widened. These findings highlight digital inclusion as a vital pathway for older adults to remain integral participants in an evolving digital society.

Rapid population aging alongside accelerating digitalization has intensified concerns about digital inequality in later life (Poole et al., 2021; Yao et al., 2021). While younger cohorts have largely integrated digital technologies into everyday life, many older adults remain digitally excluded, and inequalities within the older population have become an increasingly salient dimension of social stratification (Friemel, 2016). In China, the challenge is especially pressing (Li & Kostka, 2024). By the end of 2023, 21.1% of the population were aged 60 or older. At the same time, China continues to promote technological innovation as a driver of social and economic change, yet recent studies suggest that digital exclusion among older adults remains pronounced (Lu et al., 2022).

Digital exclusion in later life can be understood as an involuntary experience of limited engagement with the digital world and a constrained ability to benefit from it (Fang et al., 2025). Older adults are far from a monolithic group, and digital exclusion is often intertwined with broader forms of social exclusion (Helsper, 2017). In China, it is embedded in long-standing socioeconomic and regional disparities (Wang et al., 2021) and further compounded by health-related inequalities that become more prominent with age (Ding et al., 2023). These disparities are not static; they unfold across the life course and intersect with the timing of rapid digital change. A life course perspective therefore provides a dynamic framework that links age-related change to period dynamics and cohort-specific life chances (Mayer, 2009), motivating an age–period–cohort (APC) lens for distinguishing aging-related change from period shifts and cohort replacement. Examining China also provides a valuable non-Western vantage point for understanding how social structure, regional disparities, and government-led digital policies jointly shape digital inclusion, with potential implications for other rapidly aging societies.

**Urban-rural and regional disparities in digital exclusion**

Among middle-aged and older adults, digital exclusion is closely tied to socioeconomic stratification. Cumulative inequality theory posits that unequal access to social and material resources accrues across the life course and becomes especially consequential in later life (Ferraro & Shippee, 2009). Accumulated disadvantage can limit opportunities to acquire digital access and skills, whereas accumulated advantage expands the capacity to adopt and sustain engagement with digital technologies. In practice, these life-course inequalities are often institutionalized and spatially organized, most visibly through urban–rural residence and regional stratification (Scheerder et al., 2017), thereby structuring digital exclusion at older ages.

China's urban–rural dual structure provides a salient institutional context in which socioeconomic disparities translate into persistent gaps in digital access and capability. Although digital technologies can, in principle, benefit rural residents, territorial differences in infrastructure and constraints on digital literacy can impede adoption and meaningful use. Recent work suggests that the expansion of the digital economy may be associated with widening urban–rural disparities in China (Peng & Dan, 2023). Related evidence further indicates that in rural settings, limited education and economic capital can constrain the extent to which digital access is converted into broader psychosocial or economic gains (Liu et al., 2024; Zhao & Kuang, 2025).

Regional development has also unfolded unevenly across China, shaping substantial variation in Information and Communications Technology (ICT) infrastructure and diffusion. Differences in regional ICT development have been linked to unequal socioeconomic opportunities across the country (Wang et al., 2021). Consistent with this view, older adults in more digitally advanced areas, such as eastern China, appear better positioned to benefit from internet use than those in less developed regions, including parts of western China (Xue & Liu, 2024; Zhong & Wang, 2023). Taken together, digital exclusion in later life is embedded in institutional and regional structures that accumulate across the life course, motivating our focus on urban–rural and regional heterogeneity.

**Health-related factors in digital exclusion in the aging process**

A growing body of research links internet use to psychosocial and health-related outcomes in later life, including mental health (Luo et al., 2024), cognition (Kim & Han, 2022), and physical health (Cotten, 2021). Building on this literature, we shift attention to the reverse pathway by considering how health conditions in later life, particularly cognitive functioning and multimorbidity, can heighten vulnerability to digital exclusion. Health constraints may not only reduce the capacity to acquire and maintain digital skills, but also increase the effort required to navigate digital services and interfaces, thereby widening disparities in digital engagement.

For older adults, cognitive functioning is closely linked to internet use, as risks of cognitive impairment become more prevalent in later life, a concern that is gaining increasing attention in the Chinese context (Jia et al., 2020). Cognitive decline can diminish older adults' digital competence, limiting their ability to engage with digital technologies (Heponiemi et al., 2023). Meanwhile, the cognitive demands imposed by complex digital interfaces and excessive information can serve as additional barriers, even when internet access is available (Hunsaker & Hargittai, 2018). These challenges not only limit digital inclusion but may also reinforce digital exclusion. In addition, considering that many older adults suffer from at least one chronic condition, and often multimorbidity—the presence of at least two chronic diseases (Skou et al., 2022), the physical limitations associated with these health issues can further impede their ability to engage with digital technologies (Lee et al., 2011). As such, this study will also examine multimorbidity as a critical risk factor in understanding digital exclusion among older adults, acknowledging the compounded challenges that arise when cognitive and physical health decline.

**Analytical framework: APC approach**

Existing studies on digital exclusion among older adults have often relied on single-wave cross-sectional snapshots (Friemel, 2016; Lee et al., 2011; Ragnedda et al., 2022). Although such designs are informative about digital inequality at particular moments, they provide limited leverage for distinguishing aging-related change from period dynamics and cohort replacement, and therefore offer only a partial view of how digital exclusion unfolds over time. Aging is a socially embedded process shaped by both individual circumstances and broader structural contexts (Ferraro & Shippee, 2009). To move beyond one-off snapshots, we adopt an APC framework suited to repeated survey waves, which enables a dynamic, population-level account of digital exclusion across historical time. In addition, the digital exclusion literature remains disproportionately Western-focused (Kim & Han, 2022; Poole et al., 2021; van Ingen & Matzat, 2018). Even studies on China often provide limited attention to how structural inequality intersects with aging processes over time (Liu et al., 2024; Xue & Liu, 2024).

APC analysis is widely used to characterize social change and demographic dynamics by separating age, period, and cohort components (Yang & Land, 2016). The age dimension captures how later-life transitions in social roles and resources, such as retirement and health decline, are associated with digital engagement and potential withdrawal from digitally mediated activities. The period dimension reflects historical shifts that influence all age groups, including China's rapid digital expansion since 2010 alongside policy-driven digitalization of services (Yan & Schroeder, 2020). Yet period change may be unevenly experienced, and shocks such as the COVID-19 pandemic may have intensified barriers for older adults (Seifert et al., 2021). Cohort, defined by birth year as the linkage between age and historical time (Ryder, 1965), indexes the extent to which formative exposure, educational opportunities, and technology environments differ across generations (Elder et al., 2003). For instance, cohorts born in the 1930s and 1940s encountered markedly different technological conditions from those born in the 1960s and 1970s, with implications for later-life access to and attitudes toward digital tools.

**Current study**

Drawing on three nationally representative Chinese datasets, we characterize population-level age, period, and cohort patterns of digital exclusion among middle-aged and older adults. We then assess whether these patterns vary by urban–rural residence, region, and late-life health. By synthesizing evidence across surveys with diverse designs, we strengthen the robustness and generalizability of our findings in a rapidly digitalizing, non-Western context. Tracing the trajectories of digital exclusion and characterizing marginalized groups provides a critical evidence base for targeted, age-friendly inclusion policies aimed at fostering digital equity in later life.

## Methods

**Data**

Data were driven from three nationally representative surveys: China Health and Retirement Longitudinal Study (CHARLS), China Family Panel Studies (CFPS), and Chinese General Social Survey (CGSS). These surveys vary in their target populations and thematic emphasis, their collective use allows for the identification of robust and convergent patterns. Specifically, CHARLS tracks individuals aged 45 and older over time, offering rich longitudinal data on health and aging in China (Zhao et al., 2014). CFPS, another longitudinal survey, focuses primarily on households, including older adults (Xie & Hu, 2014). Its inclusion allows for an examination of whether consistent patterns emerge when examining individuals within a broader family context. CGSS is a repeated cross-sectional survey designed to capture population-level changes in social attitudes and related socio-demographic characteristics over time. To ensure temporal comparability, we aligned CGSS to the observation window of CHARLS and CFPS, using CHARLS (2011, 2013, 2015, 2018, 2020), CFPS (2010, 2014, 2016, 2018, 2020, 2022; the 2012 wave was excluded owing to the absence of digital-exclusion measures), and CGSS (2010, 2011, 2012, 2013, 2015, 2017, 2018, 2021).

We restricted the analytic sample to respondents aged 45 years and older. To mitigate data sparsity at the oldest ages, we further truncated the age range to 45–83 in CHARLS, 45–86 in CFPS, and 45–88 in CGSS. These restrictions implied corresponding birth cohort ranges of 1935–1974 for CHARLS and CFPS, and 1930–1974 for CGSS. Prior to the APC analyses, we excluded observations with missing survey weights or missing digital exclusion measures. The resulting analytic samples comprised 81,451 observations in CHARLS, 85,469 observations in CFPS, and 50,721 observations in CGSS.

**Measures**

***Digital exclusion.*** We used the variable on internet access to measure digital exclusion. In CHARLS, participants were asked, "Have you done any of these activities in the last month?" Those who selected "Used the Internet" were categorized as digitally included; those who did not were categorized as digitally excluded. In CFPS, the measurement of digital exclusion differed across waves.

From 2016 onwards, respondents were classified as digitally included if they reported using the internet via either mobile devices or a computer, and as digitally excluded if they reported neither. In the two earlier waves, digital exclusion was based on a single item on any internet use, with non-use coded as digital exclusion. In CGSS, participants were asked about the frequency of their internet use over the past year, using a 5-point scale ranging from 1 ("never") to 5 ("very frequently"). Those who selected 1 were classified as digitally excluded; all others were considered digitally included. Finally, digital exclusion was operationalized as a binary variable, where a value of 1 denoted digital exclusion and 0 denoted digital inclusion. Harmonization of digital exclusion across datasets see Table S1.

*Age and cohort.* Age in each dataset was defined as the respondent's age at the time of the survey year. The age range covered individuals from midlife (45–65) to late life (>65 years). The birth cohort was grouped into 5-year intervals and treated as a continuous variable. In CHARLS and CFPS, values ranged from 1 to 8; in CGSS, they ranged from 1 to 9.

*Subgrouping variables.* We examined heterogeneities in the association between digital exclusion and outcomes across residence, region, multimorbidity, and cognitive risk status. Residence was coded as a dichotomous indicator of urban versus rural living. Region followed the National Bureau of Statistics of China (2021) classification, distinguishing Northeast, East, Central, and West. Multimorbidity status was defined based on the number of self-reported chronic conditions. In CHARLS, fourteen chronic conditions were assessed, and respondents were categorized as having no chronic disease, one chronic disease, or multimorbidity (≥2 chronic diseases). Cognitive function was measured using a reduced form of the Telephone Interview for Cognitive Status (TICS) (Brandt et al., 1988), comprising self-rated memory, word recall, serial subtraction, date recognition, and visuospatial ability (total score range: 0–36). To identify individuals at elevated cognitive risk, we applied latent profile analysis and used a bootstrap-based cut-point procedure to determine a threshold of 18 (see Supplementary Methods). Respondents were then classified as cognitively at risk versus not at risk.

**Statistical analysis**

***Missing data and weighting***

To preserve population representativeness and mitigate bias from missing data, we combined multiple imputation for covariates with inverse probability weighting for item nonresponse in the digital exclusion measure among interviewed respondents. First, we imputed missing values in covariates using multiple imputation by chained equations (Buuren & Groothuis-Oudshoorn, 2011), generating 20 completed datasets. The digital exclusion outcome itself was not imputed.

Next, restricting to observations with a completed interview in a given wave and available cross-sectional survey weights, we constructed stabilized inverse probability weights (Cole & Hernán, 2008; Robins et al., 2000) for item nonresponse in digital exclusion. The numerator model included time-invariant predictors. The denominator model additionally included time-varying socioeconomic and health predictors and, for the longitudinal CHARLS and CFPS data, prior-wave information related to measurement availability (including an indicator for having a prior observation, whether digital exclusion was observed in the previous wave, and the lagged digital-exclusion value when available). Coefficients from these response-propensity models were pooled across imputations using Rubin's rules (Buuren & Groothuis-Oudshoorn, 2011), and predicted response probabilities were generated from the pooled models. Stabilized item-nonresponse weights were then constructed as the ratio of the predicted response probability from the numerator model to that from the denominator model, improving stability relative to unstabilized weights. Final analysis weights were obtained by multiplying these stabilized item-nonresponse weights by the wave-specific cross-sectional survey weights. The combined weights were standardized to have a mean of 1 within wave. As diagnostic, we assessed covariate balance using absolute standardized mean differences (SMD). We considered an absolute SMD below 0.25 (Stuart, 2010) to indicate acceptable balance and additionally report balance using 0.10 (Austin, 2011) as a more stringent benchmark. All weights achieved acceptable covariate balance. Detailed results are reported in the Supplementary Materials.

For CHARLS, we additionally addressed missingness in the health measures used to define subgroups. Because these subgroup analyses require non missing information on chronic conditions and cognitive function, restricting the sample to complete cases could introduce additional selection beyond item nonresponse in digital exclusion. We therefore repeated the stabilized inverse probability weighting procedure after redefining the response as the joint availability of digital exclusion and, separately, multimorbidity status or cognitive function. The resulting two sets of weights were then multiplied by the wave specific cross sectional survey weights and standardized within wave. These weights were used in the multimorbidity status stratified analyses and the cognition function stratified analyses, respectively. Details on the weights used in each model are provided in Table S2.

### *Age-period-cohort analysis*

Because age (A), period (P) and birth cohort (C) are perfectly linearly dependent (A = P − C), conventional APC specifications are not identified without additional constraints or modelling assumptions (Bell, 2020). We adopted a hierarchical APC (HAPC) specification for repeated cross-sectional data (Yang & Land, 2008) and estimated APC patterns in digital exclusion using mixed-effects logistic regression across the three datasets. Specifically, we treated age as an individual-level predictor (Yang & Land, 2008), represented period using fixed effects, and modelled cohort differences via random intercepts. Age (in years) was centered at 45 and entered as both a linear and a quadratic term to accommodate potential non-linear age gradients. Period differences were captured by including indicator variables for each survey wave, allowing estimation of period-specific deviations net of age and cohort. Cohort was operationalized as five-year birth-cohort groups and specified as a random intercept, allowing the baseline log-odds of digital exclusion to vary across cohorts conditional on age and period.

Because age was the individual-level predictor, we adjusted only for covariates that could plausibly confound the age–digital exclusion association beyond the APC components. Accordingly, we included sex to account for sex differences that may influence digital exclusion and the age composition of survey samples (Bartram, 2020). Models were estimated in Stata 18.5 using *melogit*, with probability weights given by the final analysis weights, constructed by multiplying the wave specific cross sectional survey weights with the stabilized inverse probability weights for item nonresponse. Robust standard errors were clustered by cohort group to account for potential within cohort dependence. As a sensitivity analysis, we re-estimated all models without weights to assess whether the findings were driven by the weighting strategy.

### *Bounding analysis*

To evaluate how sensitive our conclusions are to the identifying assumptions implicit in HAPC specifications, we complemented the HAPC model with a bounding analysis (Fosse & Winship, 2019b). This step is motivated by prior work showing that HAPC models can impose non-obvious constraints on the linear components of APC trends (Bell & Jones, 2018) and, in applied settings, may effectively pull the linear cohort trend toward zero (Fosse & Winship, 2019a). Rather than selecting a single identification strategy, the bounding analysis makes assumptions explicit and quantifies the range of APC decompositions consistent with both the observed data and theoretically motivated constraints (Rohrer, 2025).

We implemented the bounding analysis using the *apcR* package developed by Fosse and Winship (https://scholar.harvard.edu/apc/software-0; downloaded version June 2025). Following the linearized APC framework, we discretized the temporal dimensions into equal-width bins: age and period were grouped into two-year categories, creating a regular age-by-period grid for cell-based estimation. To maintain biennial spacing in period, we restricted the analysis to years that align with this design (CFPS: 2014–2022 at two-year intervals; CGSS: 2013–2017 at two-year intervals). We then estimated the linearized APC model on aggregated age-period-cohort cells, where cell-level digital exclusion was computed as a weighted mean using combined weights obtained by multiplying the wave-specific cross-sectional survey weights with stabilized inverse probability weights for item nonresponse. This procedure separates the overall APC pattern into (i) two identifiable linear combinations: $\theta_1 = (\alpha + \pi)$, the combined linear slope of age and period, and $\theta_2 = (\pi + \gamma)$, the combined linear slope of period and cohort; and (ii) empirically identified non-linear deviations (departures from linearity) for age, period, and cohort. Because these non-linear deviations are identified from the observed data and remain the same across all equally well-fitting APC decompositions, they provide a fixed reference for bounding the under-identified linear components.

The identification problem implies a continuum of equally well-fitting decompositions of the linear trends (the canonical solution line). To narrow this solution space, we imposed three constraints on the linear components over the observed ranges: (i) a positive linear age trend (older age associated with higher digital exclusion) (Charness & Boot, 2009), (ii) a negative linear period trend (digital exclusion declining over historical time) (Li et al., 2025), and (iii) a negative linear cohort trend (later-born cohorts exhibiting lower digital exclusion) (Matthews et al., 2019). We visualized these constraints using the two-dimensional APC bounding graph, highlighting the segments of the canonical solution line that satisfy each constraint (see Fig. S1-S6). The intersection of these admissible segments defines the bounded solution set. From this set, we derived the maximum and minimum allowable slopes for the linear age, period, and cohort components. Finally, we combined these slope bounds with the estimated non-linear deviations to reconstruct upper and lower bounded trajectories for each temporal effect. We repeated the same procedure for the full sample and for each subsample, using the bounding results as a transparent robustness check on the patterns implied by the HAPC estimates.

## Results

### CHARLS

In the CHARLS sample, digital exclusion reached 86.96%, defined as no Internet use within the past month. The sample was predominantly rural and concentrated in the Eastern region, with the 1950–1964 birth cohort constituting the majority (Table S3).

*Age effect.* Digital exclusion increased with age, with nonlinearity as indicated by the significant quadratic age term in Table 1. Predicted probabilities showed a modest downturn in midlife followed by a steady rise into later life.

***Period effect.*** Predicted digital exclusion declined modestly from 2011 to 2018 and then dropped sharply by 2020 (Fig. 1).

***Cohort effect.*** Cohort differences were most pronounced for those born in the late 1940s and early 1950s, who showed positive cohort-specific intercept deviations. By contrast, cohorts born from the mid-1960s onward exhibited consistently negative deviations, indicating lower baseline digital exclusion net of age and period.

***Urban–rural differences.*** Predicted probabilities for rural residents consistently exceeded those for urban residents across all ages. While both trajectories displayed a modest decline in early midlife followed by a steady rise, the disparity between the two groups narrowed with advancing age. Period-specific predictions declined in both groups, yet rural levels remained higher throughout, with no obvious reduction in the urban-rural gap over time. Cohort variation was comparatively limited in the urban populations.

***Regional differences.*** Predicted digital exclusion was highest in the West. Regional gaps narrowed with increasing age, while the East showed the most pronounced cohort differentiation.

***Health-related differences.*** Stratifying CHARLS by multimorbidity status indicated the lowest predicted digital exclusion among those without chronic disease and the highest among those with multimorbidity; separation was largest in midlife and smaller at older ages as all curves approached similarly high levels. Cohort deviations were more dispersed in the no-chronic-disease group, whereas later-born cohorts tended to show lower digital exclusion within the multimorbidity group. Regarding cognitive function, the not-at-risk group exhibited a non-linear age profile characterized by a modest midlife decline followed by a gradual increase. In contrast, the at-risk group remained consistently higher at every age, with the disparity between the two groups being most pronounced between ages 50 and 70. Cohort deviations declined steadily in the not-at-risk group, whereas cohort differences were smaller and less systematic in the at-risk group. While period-specific predictions declined in both subgroups, the gap between the risk and non-risk groups gradually widened over time. (Table 4; Fig. 2).

**CFPS**

In the CFPS dataset, 78.61% of observations were digitally excluded based on the "any internet use" measure, with the analytic sample skewed slightly toward rural residents (52.82%) and the 1950–1969 birth cohorts.

***Age effect.*** Following a slight decline in the late 40s, digital exclusion increased steadily with age. This upward trend was most pronounced between ages 50 and 70, after which the trajectory plateaued at advanced ages and showed a marginal downturn among the oldest-old.

***Period effect.*** A shift in the period effect emerged after 2014, as the trend accelerated from a slow decline to a precipitous drop that hit its minimum in 2022 (Fig. 3; Table 2).

***Cohort effect.*** Cohort difference was comparatively modest, with most cohort-specific intercept deviations clustering close to zero. One exception was the 1940–1944 cohort, which exhibited a clearly positive deviation.

***Urban–rural differences.*** Predicted probabilities remained higher in rural compared to urban populations across all ages, with rural group displaying a non-linear age shape (Table 2). The period-based decline was observed in both settings but was steeper among urban populations. Cohort deviations further differentiated the two groups: the urban populations showed lower exclusion for 1945–1954 and higher levels for 1940–1944 and 1965–1969. In comparison, the rural populations were characterized by lower deviations for the 1955–1969 cohorts and elevated risks for the 1940–1944 and 1970–1974 groups.

***Regional differences.*** The West tended to remain at higher predicted levels, whereas regional differences narrowed at older ages as trajectories converged. In the East, cohorts born in 1940–1944 and 1965–1969 showed elevated digital exclusion relative to adjacent cohorts.

**CGSS**

CGSS recorded a digital exclusion rate of 69.20%, measured via internet use frequency over the preceding year. Demographics showed an urban majority (58.59%) and a regional distribution concentrated in the East (36.86%).

***Age effect.*** Predicted probabilities rose steadily from midlife into later life and then approached a plateau at the oldest ages.

***Period effect.*** The downward trajectory was evident across the full observation window and became steeper after 2015.

***Cohort effect.*** Earlier cohorts showed negative deviations, followed by positive deviations among cohorts born from the late 1940s through the 1960s. The most recent cohort group returned closer to zero.

***Urban–rural differences.*** Rural populations consistently had higher predicted digital exclusion than urban populations across ages and periods. The gap was widest in midlife and narrowed somewhat in later life, as the urban age trajectory increased more sharply with age while the rural trajectory approached a ceiling. Cohort deviations also differed by residence, with clearer cohort differentiation in urban populations and smaller deviations in rural populations (see Fig 4f).

***Regional differences.*** Predicted digital exclusion was lowest in the East. Central and Northeast China showed more clearly nonlinear age profiles, as reported in Table 3, but age trajectories across all regions became more similar at older ages. Cohort deviations also varied by region, with comparatively small cohort differentiation in Central and Northeast China.

**Robustness checks**

To assess the sensitivity of the APC findings to identification assumptions, we complemented the HAPC models with a bounding analysis in CFPS and CGSS. The results reproduced the main age and period patterns in both datasets. In CFPS, the bounded age profile showed a modest midlife dip followed by increasing risk at older ages (Fig. 5), whereas in CGSS the bounded age profile was largely monotonic with wider uncertainty at the age extremes (Fig. 6). In both surveys, period trend remained downward over time. Cohort patterns from the bounding analysis suggested an overall later-cohort advantage, characterized by a sharper decline among cohorts born after 1960. However, a distinct peak in digital exclusion was observed for those born between 1950 and the mid-1950s relative to adjacent cohorts. Overall, cohort heterogeneity remained less uniform than the observed age and period associations. In sum, the bounding analysis supports the robustness of the age and period conclusions under alternative identification assumptions and provides a plausible range for cohort variation. We also repeated all HAPC analyses without weights, and our findings remained robust (see Tables S4-S7).

**Discussion**

Across three nationally representative Chinese datasets, we identified a consistent APC pattern in digital exclusion among adults aged 45 years and older. Digital exclusion increased with age across the observed age range, whereas period estimates declined across survey years. Cohort differences were also evident but less uniform than the age and period components, with no single monotonic cohort pattern emerging across datasets. Structural disparities by residence and region were more pronounced at younger old ages and tended to narrow at the oldest ages. These core conclusions were reinforced by robustness checks, including bounding analyses and re-estimation without weights, which reproduced the main age and period patterns and delineated a plausible range for cohort variation.

Age exhibited a robust positive association with digital exclusion independent of period and cohort effects. Beyond this overall trend, the age trajectory was modestly non-linear, showing a mild attenuation around midlife followed by a steadier increase at older ages. This midlife easing may reflect relatively stronger resources and capacities (Infurna et al., 2020; Lachman et al., 2015) that help sustain routine digital contact and delay disengagement. At older ages, declines in physical and cognitive functioning (Czaja et al., 2006) may raise the costs of learning, troubleshooting, and maintaining digitally mediated activities, potentially offsetting perceived usefulness (Davis, 1989) and contributing to higher digital exclusion risk. Life-course transitions may further reinforce this pattern (Faure et al., 2020): retirement and shifts in household or family roles can reduce everyday exposure to work-based digital environments and digital tasks, weakening opportunities for skill maintenance and habit formation. More broadly, role-transition perspectives in gerontology (Rosow et al., 1976) suggest that changes in later-life role structures can reorganize daily routines and, in turn, reshape opportunities for digital engagement.

Period effects reveal a decline in digital exclusion across the 2010s and early 2020s, in line with prior research showing that population-level access to digital technologies has expanded over time (Ren & Zhu, 2024; Zhao & Kuang, 2025). This downward period pattern is consistent with widening connectivity and the diffusion of digitally mediated services (CNNIC, 2022), although the pace of improvement is not uniform across survey windows. Over the past decade, China has steadily advanced digitalization through investments in internet infrastructure and the rapid scaling of everyday digital ecosystems, including mobile payments and remote health services, under the broader policy agenda of "Digital China" (Ito, 2019). The COVID-19 pandemic further accelerated these shifts by normalizing digital governance and tracking technologies, such as health codes (Li & Kostka, 2024). Together, these period-specific developments likely exerted simultaneous influence across age groups and birth cohorts, shifting the overall risk of digital exclusion downward over historical time while leaving room for persistent stratification.

Although cohort differences were evident, they were less consistent across the three datasets than the age and period effects. The 1930–1974 birth cohorts experienced a major shift from predominantly analog media to the early diffusion of the internet; earlier cohorts were less likely to be exposed to the internet during formative stages of the life course, whereas later-born cohorts—especially those reaching early adulthood around the turn of the 21st century—could engage with digital technologies earlier and adopt them more rapidly. In China, this generational positioning also coincided with post-reform expansions in education and changing institutional demands, plausibly supporting greater skill formation and capacity for digital engagement among later cohorts.

However, a notable exception is that cohorts born in the early- to mid-1950s exhibit a local peak in digital exclusion. This pattern likely reflects the distinctive socio-historical experiences of this generation. On the one hand, these cohorts were raised during periods of political upheavals and material scarcity, including the Great Chinese Famine and the Cultural Revolution, which disrupted everyday life and institutional environments during childhood and adolescence (Chen et al., 2020; Zhou & Hou, 1999). During the critical developmental window of ages 16 to 20, their educational opportunities were severely disrupted, leading to gaps in attainment that were rarely recovered in subsequent decades (Deng & Treiman, 1997). On the other hand, empirical evidence suggests that these cohorts may exhibit diminished cognitive functioning compared to adjacent cohorts (Guo & Zheng, 2023). Given that education is a primary predictor of internet adoption (Lythreatis et al., 2022; van Ingen & Matzat, 2018)—and that cognitive function are essential for maintaining digital activities—the combination of early educational deprivation and later cognitive declines explains why the 1950s cohorts face a particularly elevated risk of digital exclusion in later life.

Digital exclusion is embedded in offline circumstances and interacts with social inequalities, linking digital disadvantage to broader processes of social exclusion (Helsper, 2012; Ragnedda et al., 2022). In our analyses, structural disparities were pronounced across both residence and region. Rural residents consistently showed higher predicted levels of digital exclusion than their urban counterparts. This persistent gap aligns with the framework of cumulative inequality (Ferraro & Shippee, 2009), suggesting that early-life or structural deficits in resource access create long-term trajectories of disadvantage. Despite China's efforts to promote rural digitalization, rural areas continue to lag behind cities in the broader ecosystem that supports digital inclusion. Regional stratification followed a similar pattern: digital exclusion tended to be lower in the East and higher in the West, aligning with evidence that reductions in digital divides have not been evenly distributed across geographic contexts (Li et al., 2025). Such patterns reflect the self-reinforcing logic of cumulative advantage (Dannefer, 1987), manifesting as a Matthew Effect within the digital landscape. Resource-rich regions are better positioned to mobilize new technologies to consolidate their socio-economic standing, thereby widening the gap with historically disadvantaged areas. These spatial disparities persisted across the decade-long study period, indicating that the inclusive expansion of digital technologies has not sufficed to dismantle entrenched social inequalities (Witte & Mannon, 2010).

However, residence- and region-based gaps converged among the oldest age groups. One plausible interpretation is that, at advanced ages, ubiquitous constraints related to health, functional limitations may override spatial advantages, compressing between-group differences—a pattern often framed in life-course debates as the "age-as-leveler" hypothesis (Lynch, 2003).

Digital exclusion is not only socially patterned but also closely intertwined with late-life health vulnerability. Respondents with multimorbidity exhibited markedly higher predicted exclusion, corroborating prior research that health-related difficulties constitute significant barriers to internet engagement among older adults (Ang et al., 2021). Multimorbidity may trigger loss-based selection in later life: as illness management becomes the dominant goal, older adults often narrow their activity repertoires to conserve finite energy and attention. From the Selection, Optimization, and Compensation perspective (Baltes, 1997), reduced digital engagement arises because maintaining competence in digital environments demands continuous optimization and compensation—an investment of adaptive resources that becomes arguably unsustainable when the burden of illness is high. Particularly, we found that disparities based on multimorbidity status were widest prior to age 60 and diminished significantly with age. This narrowing gap reinforces the argument that late-life health and functional constraints may act as a dominant leveler. Ultimately, the high adaptive costs of digital technology may create a ceiling effect in advanced age that can erode the comparative advantage of even the healthier older adults.

Cognitive risk also stratified digital exclusion in later life, with elevated exclusion evident already in younger old age and remaining high thereafter, alongside weaker cohort differentiation—patterns suggesting that cognitive constraints may blunt whether later-born cohorts' greater digital exposure converts into actual technology use. In line with this interpretation, some research (Dequanter et al., 2022; Heponiemi et al., 2023) highlights cognitive functioning as a key determinant of technology use, helping explain why some older adults may struggle to turn exposure opportunities into uptake. Admittedly, China's rapid infrastructure expansion may have lowered structural barriers to basic internet connectivity (World Bank, 2026). Yet our results suggest that the cognitive-risk gap did not narrow over period and may have widened, consistent with the possibility that, even as access becomes easier, the cognitive demands of use become more salient as digital systems grow more complex and require ongoing adaptation (Li & Luximon, 2020).

Our study has several strengths. First, by applying an APC framework to three nationally representative surveys with distinct designs, we assess whether key patterns replicate across measurement windows and sampling structures, strengthening inference beyond any single dataset. Second, we acknowledge debates about HAPC identification (Bell, 2020; Bell & Jones, 2018) and therefore complement the HAPC models with bounding analyses as a robustness check. Third, whereas much recent work quantifies the digital divide through dimensions of access or use (Ren & Zhu, 2024; Zhao & Kuang, 2025), our analyses emphasize how digital exclusion shifts across historical time and aging, and identify which underlying forms of inequality continue to structure exclusion even as overall digital environments expand. Consistent with this aim, we document how digital exclusion is patterned by residence, region, and late-life health vulnerability, showing that population-level improvements can coexist with persistent subgroup disparities, particularly when multimorbidity and cognitive risk raise the barriers to sustaining digital engagement.

Several limitations warrant consideration. First, although we harmonized the digital-exclusion measure across surveys, it was operationalized with different reference periods and question formats. While all three datasets are nationally representative, they differ in design and substantive focus: CHARLS is an aging-focused panel, CFPS is a household-based panel, and CGSS is a general social survey with a comparatively smaller older-adult sample. Such differences in measurement and sample composition may affect cross-survey comparability, including the apparent timing of period shifts and the fine-grained shape of estimated cohort deviations. Second, our binary indicator captures only a basic threshold of digital exclusion and does not account for digital skills, quality of access, or the capacity to effectively use and benefit from digital services. Future studies could extend the APC framework by incorporating multidimensional measures of digital inclusion and by examining outcomes that more directly reflect the consequences of exclusion.

## Conclusion

Drawing on three nationally representative surveys, this study provides a comprehensive assessment of the dynamics of digital exclusion among middle-aged and older adults in China. Analysis grounded in the APC framework reveals that despite rapid digitalization over historical time and later-cohort advantages in digital literacy, aging itself remains a persistent barrier to digital participation. We also identify a distinct excess risk concentrated among cohorts born in the early- to mid-1950s. These findings illuminate how digital exclusion is inscribed within the life course, deepening as individuals navigate the physiological vulnerabilities and social marginalization that accompany advanced age. Moreover, the results underscore that digital exclusion in later life is closely intertwined with broader structures of social inequality. Despite continuous improvements in national digital infrastructure, exclusion remains systematically patterned by urban–rural residence, regions, and health status, with particularly elevated risks among those living with multimorbidity and at higher risk of cognitive impairment. Infrastructure expansion and technological diffusion alone are therefore unlikely to achieve digital equity. Consequently, advancing digital inclusion requires targeted, age-friendly policies that go beyond access alone. Policy efforts should adopt a life-course perspective that recognizes cohort-specific constraints on digital uptake, prioritize sustained and locally tailored support for groups facing cumulative disadvantage, and reduce cognitive and physical barriers through accessible, low-burden design and community-embedded assistance to prevent digitalization from compounding late-life vulnerabilities as population aging accelerates.

**Transparency and Openness**

Datasets are available from the China Health and Retirement Longitudinal Study (CHARLS) at https://charls.charlsdata.com, the Chinese General Social Survey (CGSS) at https://www.cnsda.org, and the China Family Panel Studies (CFPS) at https://www.isss.pku.edu.cn/sjsj/cfpsxm/index.htm. Access to these datasets is available upon registration and application. The analysis code is available at https://osf.io/qd7s6/.

Table 1 Weighted Multilevel Logistic Regression of Digital Exclusion in CHARLS

| | Full sample | Urban | Rural | East | Central | West | Northeast |
|---|---|---|---|---|---|---|---|
| Intercept | 8.863 [5.210,15.078]*** | 2.131 [1.641,2.766]*** | 43.443 [32.932,57.308]*** | 5.405 [3.331,8.772]*** | 9.143 [7.423,11.262] *** | 10.787 [9.438,12.329]*** | 5.830 [3.417,9.947]*** |
| Age | 1.040 [0.993,1,089] | 1.071 [1.052,1.090]*** | 1.177 [1.128,1.228]*** | 1.041 [0.985,1.100] | 1.068 [1.005,1.136]* | 1.125 [1.094,1.157]*** | 1.089 [1.010,1.173]* |
| Age$^2$ | 1.001 [1.000,1.002]** | 1.001 [1.000,1.001]** | 1.000 [0.998,1.001] | 1.001 [0.999,1.004] | 1.001 [0.999,1.003] | 1.000 [0.998,1.001] | 1.001 [0.999,1.002] |
| Period (ref: 2011) | | | | | | | |
| 2013 | 0.594 [0.462,0.764]*** | 0.616 [0.509,0.747]*** | 0.255 [0.141,0.460]*** | 0.570 [0.392,0.828]** | 0.531 [0.397,0.710]*** | 0.615 [0.518,0.730]*** | 0.621 [0.389,0.991]* |
| 2015 | 0.366 [0.303,0.442]*** | 0.381 [0.309,0.470]*** | 0.088 [0.070,0.111]*** | 0.361 [0.298,0.437]*** | 0.307 [0.242,0.388]*** | 0.330 [0.232,0.470]*** | 0.330 [0.240,0.454]*** |
| 2018 | 0.177 [0.126,0.248]*** | 0.189 [0.156,0.229]*** | 0.024 [0.016,0.035]*** | 0.183 [0.126,0.267]*** | 0.132 [0.094,0.185]*** | 0.122 [0.105,0.142]*** | 0.136 [0.094,0.197]*** |
| 2020 | 0.035 [0.023,0.054]*** | 0.049 [0.040,0.059]*** | 0.003 [0.002,0.004]*** | 0.045 [0.030,0.068]*** | 0.023 [0.017,0.029]*** | 0.020 [0.017,0.024]*** | 0.024 [0.016,0.036]*** |
| Sex (Female = 1) | 1.670 [1.522,1.833]*** | 1.475 [1.239,1.756]*** | 2.034 [1.855,2.230]*** | 1.925 [1.467,2.526]*** | 1.550 [1.367,1.757]*** | 1.619 [1.454,1.803]*** | 1.197 [0.977,1.466] |
| *Random effects* | | | | | | | |
| σ$^2$ Cohort | 0.056 [0.002,1.489] | 0.003 [0.001,0.012] | 0.081 [0.006,1.182] | 0.072 [0.021,0.253] | 0.008 [0.005,0.106] | 0.000 [0.000,0.000] | 0.020 [0.009,0.046] |
| BIC | 55823.3 | 24583.2 | 23063.7 | 22541.7 | 13702.6 | 13678.5 | 5109.7 |
| AIC | 55758.1 | 24520.6 | 23000.3 | 22484.7 | 13646.2 | 13621.2 | 5056.4 |
| N | 81,451 | 18,502 | 62,949 | 25,686 | 23,243 | 26,681 | 5,785 |

*Note.* Age was centered at 45. Fixed-effect estimates are reported as odds ratios (ORs) with 95% confidence intervals. Models are weighted, and standard errors are clustered

at the cohort level. AIC = Akaike information criterion; BIC = Bayesian information criterion. *** p < .001, ** p < .01, * p < .05.

Table 2 Weighted Multilevel Logistic Regression of Digital Exclusion in CFPS

| | Full sample | Urban | Rural | East | Central | West | Northeast |
|---|---|---|---|---|---|---|---|
| Intercept | 6.490 | 3.402 | 27.114 | 5.310 | 5.155 | 11.818 | 5.603 |
| | [5.228,8.061]*** | [2.659,4.354]*** | [16.446,44.704]*** | [4.176,6.752]*** | [4.454,5.967]*** | [9.416,14.833]*** | [3.972,7.904]*** |
| Age | 1.130 | 1.128 | 1.198 | 1.162 | 1.103 | 1.132 | 1.150 |
| | [1.098,1.164]*** | [1.086,1.172]*** | [1.146,1.252]*** | [1.096,1.233]*** | [1.077,1.128]*** | [1.099,1.166]*** | [1.134,1.166]*** |
| Age$^2$ | 0.999 [0.998,1.000] | 0.999 [0.998,1.000] | 0.998 | 0.999 [0.997,1.000] | 1.000 [0.999,1.001] | 0.999 [0.998,1.000] | 0.999 |
| | | | [0.997,0.999]*** | | | | [0.998,0.999]*** |
| Period (ref: 2010) | | | | | | | |
| 2014 | 0.475 | 0.561 | 0.311 | 0.433 | 0.519 | 0.476 | 0.457 |
| | [0.416,0.542]*** | [0.471,0.669]*** | [0.248,0.391]*** | [0.329,0.569]*** | [0.444,0.607]*** | [0.363,0.624]*** | [0.345,0.605]*** |
| 2016 | 0.171 | 0.232 | 0.073 | 0.140 | 0.196 | 0.156 | 0.195 |
| | [0.145,0.202]*** | [0.178,0.304]*** | [0.056,0.096]*** | [0.097,0.202]*** | [0.171,0.225]*** | [0.110,0.219]*** | [0.149,0.256]*** |
| 2018 | 0.088 | 0.115 | 0.027 | 0.076 | 0.116 | 0.066 | 0.076 |
| | [0.072,0.107]*** | [0.090,0.147]*** | [0.021,0.034]*** | [0.049,0.117]*** | [0.099,0.136]*** | [0.049,0.088]*** | [0.066,0.089]*** |
| 2020 | 0.047 | 0.060 | 0.012 | 0.039 | 0.067 | 0.035 | 0.033 |
| | [0.037,0.059]*** | [0.044,0.083]*** | [0.008,0.018]*** | [0.024,0.066]*** | [0.058,0.077]*** | [0.025,0.050]*** | [0.022,0.051]*** |
| 2022 | 0.027 | 0.039 | 0.006 | 0.020 | 0.040 | 0.020 | 0.025 |
| | [0.022,0.032]*** | [0.029,0.053]*** | [0.004,0.009]*** | [0.011,0.035]*** | [0.033,0.049]*** | [0.014,0.028]*** | [0.019,0.034]*** |
| Sex (Female = 1) | 1.436 | 1.364 | 1.694 | 1.658 | 1.400 | 1.592 | 0.972 [0.731,1.292] |
| | [1.334,1.546]*** | [1.227,1.517]*** | [1.477,1.943]*** | [1.412,1.946]*** | [1.234,1.588]*** | [1.371,1.849]*** | |
| *Random effects* | | | | | | | |
| $\sigma^2$ Cohort | 0.017 [0.002,0.151] | 0.019 [0.004,0.088] | 0.042 [0.003,0.552] | 0.045 [0.003,0.577] | 0.000 [0.000,0.000] | 0.033 [0.010,0.112] | 0.007 [0.005,0.095] |
| BIC | 70706.2 | 45471.8 | 21636.1 | 25410.2 | 20150.4 | 14691.5 | 9800.8 |
| AIC | 70631.4 | 45403.0 | 21575.1 | 25352.5 | 20094.7 | 14635.1 | 9740.9 |
| N | 85,469 | 40,322 | 45,147 | 28,127 | 20,897 | 23,254 | 13,191 |

*Note.* Age was centered at 45. Fixed-effect estimates are reported as odds ratios (ORs) with 95% confidence intervals. Models are weighted, and standard errors are clustered at the cohort level. AIC = Akaike information criterion; BIC = Bayesian information criterion. *** p < .001, ** p < .01, * p < .05.

Table 3 Weighted Multilevel Logistic Regression of Digital Exclusion in CGSS

| | Full sample | Urban | Rural | East | Central | West | Northeast |
|---|---|---|---|---|---|---|---|
| Intercept | 1.532 | 0.0701 | 11.559 | 0.817 | 2.333 | 4.914 | 1.844 |
| | [0.945,2.484] | [0.478,1.026] | [7.040,18.979]*** | [0.582,1.147] | [2.061,2.641]*** | [3.044,7.912]*** | [1.171,2.905]** |
| Age | 1.090 | 1.100 | 1.139 | 1.085 | 1.159 | 1.097 | 1.136 |
| | [1.057,1.124]*** | [1.077,1.123]*** | [1.094,1.187]*** | [1.060,1.111]*** | [1.131,1.188]*** | [1.057,1.138]*** | [1.098,1.175]*** |
| Age$^2$ | 1.000 | 1.000 | 0.999 | 1.000 | 0.999 | 1.000 | 0.999 |
| | [1.000,1.001] | [1.000,1.001] | [0.998,1.001] | [1.000,1.001] | [0.998,0.999]*** | [0.999,1.001] | [0.998,1.000]* |
| Period (ref: 2010) | | | | | | | |
| 2011 | 0.836 | 0.898 | 0.624 | 0.915 | 0.653 | 0.821 | 0.882 |
| | [0.708,0.986]* | [0.731,1.103] | [0.469,0.830]** | [0.668,1.254] | [0.439,0.970]* | [0.553,1.219] | [0.641,1.214] |
| 2012 | 0.639 | 0.673 | 0.430 | 0.763 | 0.682 | 0.500 | 0.534 |
| | [0.567,0.720]*** | [0.604,0.750]*** | [0.261,0.708]*** | [0.679,0.858]*** | [0.475,0.980]* | [0.374,0.668]*** | [0.361,0.791]** |
| 2013 | 0.555 | 0.643 | 0.294 | 0.625 | 0.536 | 0.508 | 0.583 |
| | [0.467,0.660]*** | [0.548,0.755]*** | [0.195,0.443]*** | [0.495,0.791]*** | [0.444,0.646]*** | [0.393,0.658]*** | [0.354,0.961]* |
| 2015 | 0.363 | 0.377 | 0.174 | 0.395 | 0.350 | 0.247 | 0.446 |
| | [0.299,0.442]*** | [0.322,0.441]*** | [0.114,0.265]*** | [0.351,0.444]*** | [0.277,0.444]*** | [0.181,0.336]*** | [0.328,0.608]*** |
| 2017 | 0.181 | 0.194 | 0.069 | 0.216 | 0.148 | 0.129 | 0.204 |
| | [0.131,0.250]*** | [0.154,0.243]*** | [0.043,0.110]*** | [0.191,0.245]*** | [0.132,0.167]*** | [0.105,0.159]*** | [0.135,0.308]*** |
| 2018 | 0.123 | 0.161 | 0.042 | 0.154 | 0.103 | 0.080 | 0.124 |
| | [0.087,0.174]*** | [0.130,0.200]*** | [0.025,0.069]*** | [0.128,0.186]*** | [0.089,0.120]*** | [0.065,0.098]*** | [0.085,0.181]*** |
| 2021 | 0.055 | 0.088 | 0.010 | 0.098 | 0.035 | 0.024 | 0.032 |
| | [0.030,0.101]*** | [0.057,0.136]*** | [0.006,0.017]*** | [0.076,0.128]*** | [0.027,0.045]*** | [0.016,0.036]*** | [0.017,0.062]*** |
| Sex (Female = 1) | 1.305 | 1.362 | 1.470 | 1.297 | 1.696 | 1.259 | 1.057 |
| | [1.237,1.377]*** | [1.255,1.479]*** | [1.324,1.633]*** | [1.185,1.419]*** | [1.531,1.878]*** | [1.089,1.457]** | [0.886,1.261] |
| *Random effects* | | | | | | | |
| σ² Cohort | 0.115 | 0.061 | 0.002 | 0.023 | 0.000 | 0.077 | 0.002 |
| | [0.006,2.290] | [0.007,0.498] | [0.000,0.149] | [0.005,0.108] | [0.000,0.000] | [0.026,0.229] | [0.000,800.644] |
| BIC | 51877.2 | 33274.7 | 12622.9 | 21129.9 | 11639.2 | 10018.9 | 5940.7 |
| AIC | 51797.7 | 33200.0 | 12559.3 | 21059.3 | 11579.2 | 9952.0 | 5887.0 |
| N | 50,721 | 29,715 | 21,006 | 18,698 | 13,479 | 12,451 | 6,093 |

*Note.* Age was centered at 45. Fixed-effect estimates are reported as odds ratios (ORs) with 95% confidence intervals. Models are weighted, and standard errors are clustered at the cohort level. AIC = Akaike information criterion; BIC = Bayesian information criterion. *** p < .001, ** p < .01, * p < .05.

Table 4 Weighted Multilevel Logistic Regression of Digital Exclusion in CHARLS Stratified by Health Status

| | No chronic disease | Single chronic disease | Multimorbidity | Cognitively not at risk | Cognitively at risk |
|---|---|---|---|---|---|
| Intercept | 7.528 [6.030,9.397]*** | 7.157 [5.346,9.581]*** | 8.748 [5.229,14.635]*** | 5.779 [4.522,7.385]*** | 30.397 [4.494,205.607]*** |
| Age | 1.081 [1.034,1.132]*** | 1.097 [1.042,1.154]*** | 1.030 [0.924,1.148] | 1.028 [0.987,1.071] | 1.014 [0.897,1.147] |
| Age$^2$ | 1.001 [1.000,1.002] | 1.000 [0.999,1.002] | 1.001 [0.999,1.004] | 1.001 [1.000,1.003]* | 1.001 [0.998,1.004] |
| Period (ref: 2011) | | | | | |
| 2013 | 0.518 [0.347,0.772]** | 0.630 [0.458,0.867]** | 0.593 [0.461,0.763] | 0.516 [0.395,0.675]*** | 0.882 [0.519,1.499] |
| 2015 | 0.263 [0.193,0.358]*** | 0.350 [0.249,0.491]*** | 0.459 [0.319,0.661] | 0.361 [0.280,0.466]*** | 0.440 [0.185,1.046] |
| 2018 | 0.126 [0.082,0.192]*** | 0.171 [0.129,0.226]*** | 0.180 [0.119,0.272] | 0.159 [0.112,0.224]*** | 0.193 [0.060,0.619]** |
| 2020 | 0.025 [0.017,0.039]*** | 0.029 [0.021,0.039]*** | 0.039 [0.029,0.052] | 0.040 [0.029,0.057]*** | 0.040 [0.012,0.132]*** |
| Sex (Female = 1) | 1.593 [1.523,1.666]*** | 1.517 [1.184,1.944]*** | 1.918 [1.748,2.103] | 1.222 [1.045,1.429]* | 2.313 [1.762,3.036]*** |
| *Random effects* | | | | | |
| $\sigma^2$ Cohort | 0.029 [0.005,0.181] | 0.006 [0.010,0.034] | 0.010 [0.000,4.361] | 0.045 [0.011,0.196] | 0.624 [0.012,33.787] |
| BIC | 23412.9 | 13880.7 | 18489.5 | 33306.9 | 11038.5 |
| AIC | 23355.2 | 13817.1 | 18430.7 | 33248.1 | 10980.5 |
| N | 28,074 | 20,956 | 32,421 | 33,062 | 29,293 |

*Note.* Age was centered at 45. Fixed-effect estimates are reported as odds ratios (ORs) with 95% confidence intervals. Models are weighted, and standard errors are clustered at the cohort level. Multimorbidity was defined as having two or more chronic conditions. Cognitive function was measured as a composite score (range: 1–36); a cutoff of 18 was used to classify respondents as at risk of cognitive impairment versus not at risk. AIC = Akaike information criterion, BIC = Bayesian information criterion. *** p < .001, ** p < .01, * p < .05.

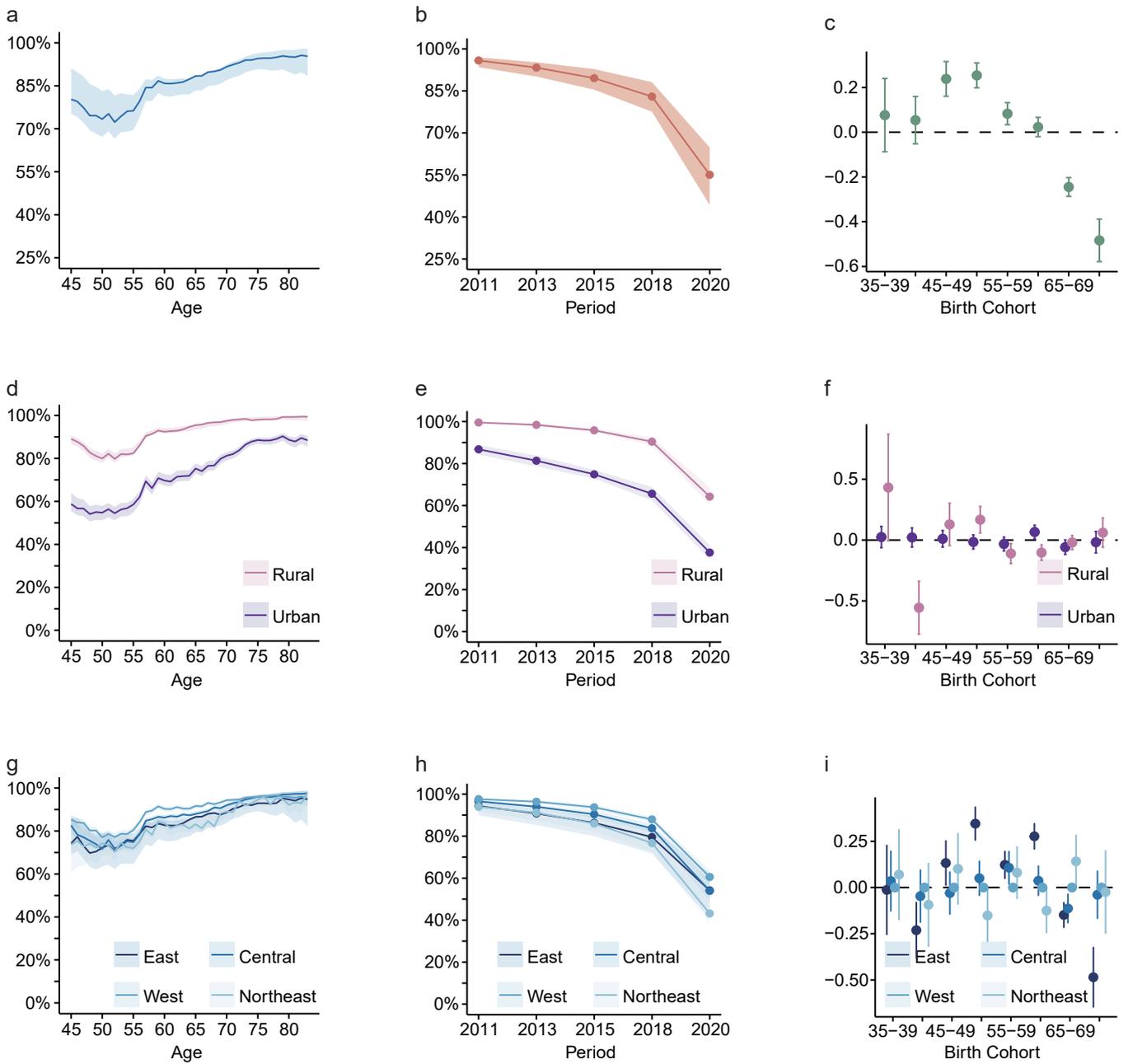

**Fig. 1 | Age–period–cohort patterns of digital exclusion in CHARLS (2011–2020).**
(a) age trajectory, (b) period trend, (c) cohort deviations. (d–f) Stratified by urban–rural residence.
(g–i) Stratified by region. In the age and period panels, the y-axis indicates the predicted probability
of digital exclusion from weighted HAPC logistic models; shaded bands denote 95% confidence
intervals. Cohort panels display the estimated random intercepts for each five-year birth cohort (on
the log-odds scale). Birth cohorts are abbreviated by the last two digits (e.g., 39 refers to 1939). The
reference line at 0 represents no deviation from the overall intercept, with positive values indicating
higher risk of digital exclusion.

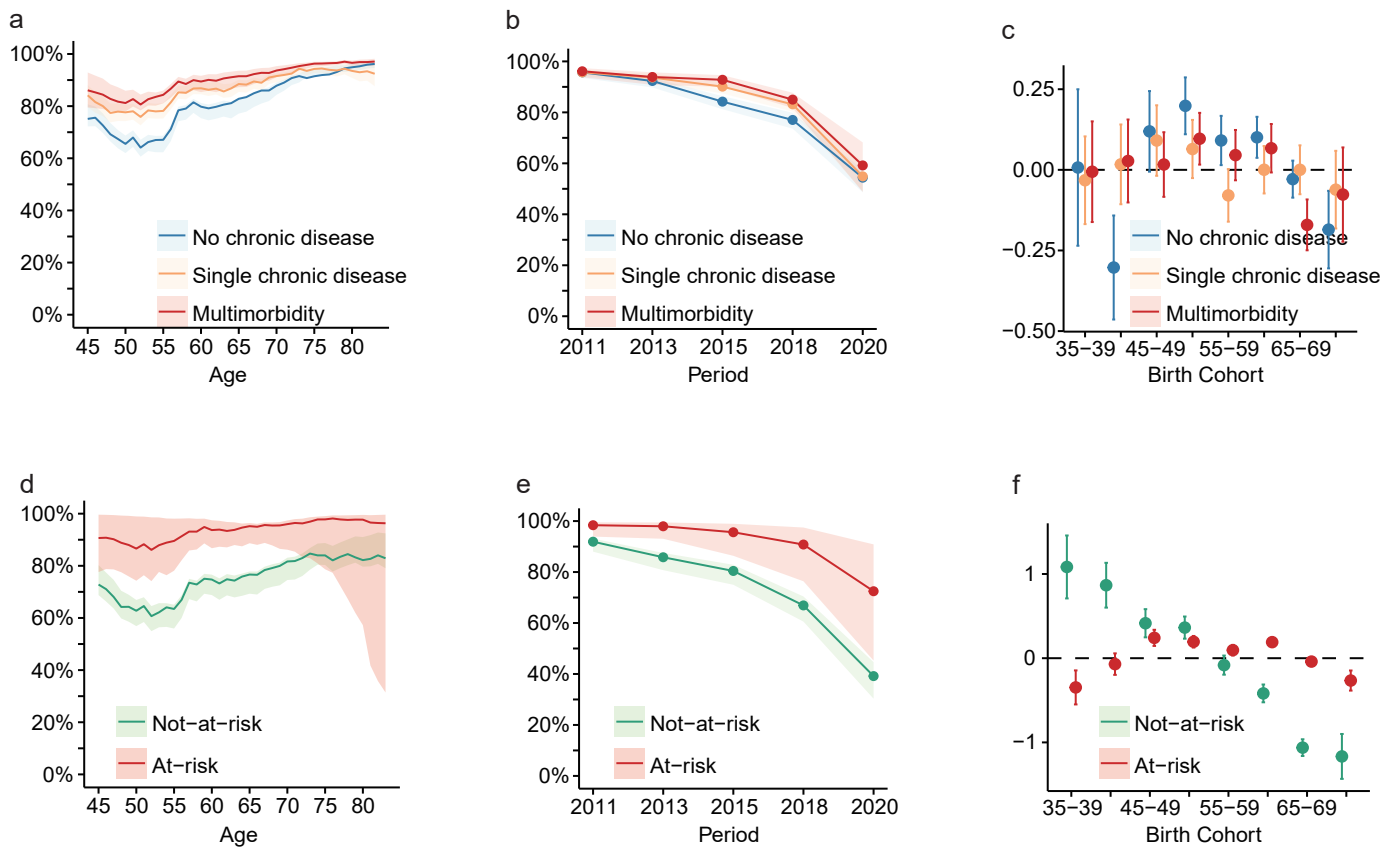

**Fig. 2 | Age–period–cohort patterns of digital exclusion in CHARLS (2011–2020), stratified by health status.**
(a–c) Stratified by multimorbidity status: (a) age trajectory, (b) period trend, and (c) cohort deviations. (d–f) Stratified by cognitive impairment risk: (d) age trajectory, (e) period trend, and (f) cohort deviations. In the age and period panels, the y-axis indicates the predicted probability of digital exclusion from weighted HAPC logistic models; shaded bands denote 95% confidence intervals. Cohort panels display the estimated random intercepts for each five-year birth cohort (on the log-odds scale). Birth cohorts are abbreviated by the last two digits (e.g., 39 refers to 1939). The reference line at 0 represents no deviation from the overall intercept, with positive values indicating higher risk of digital exclusion.

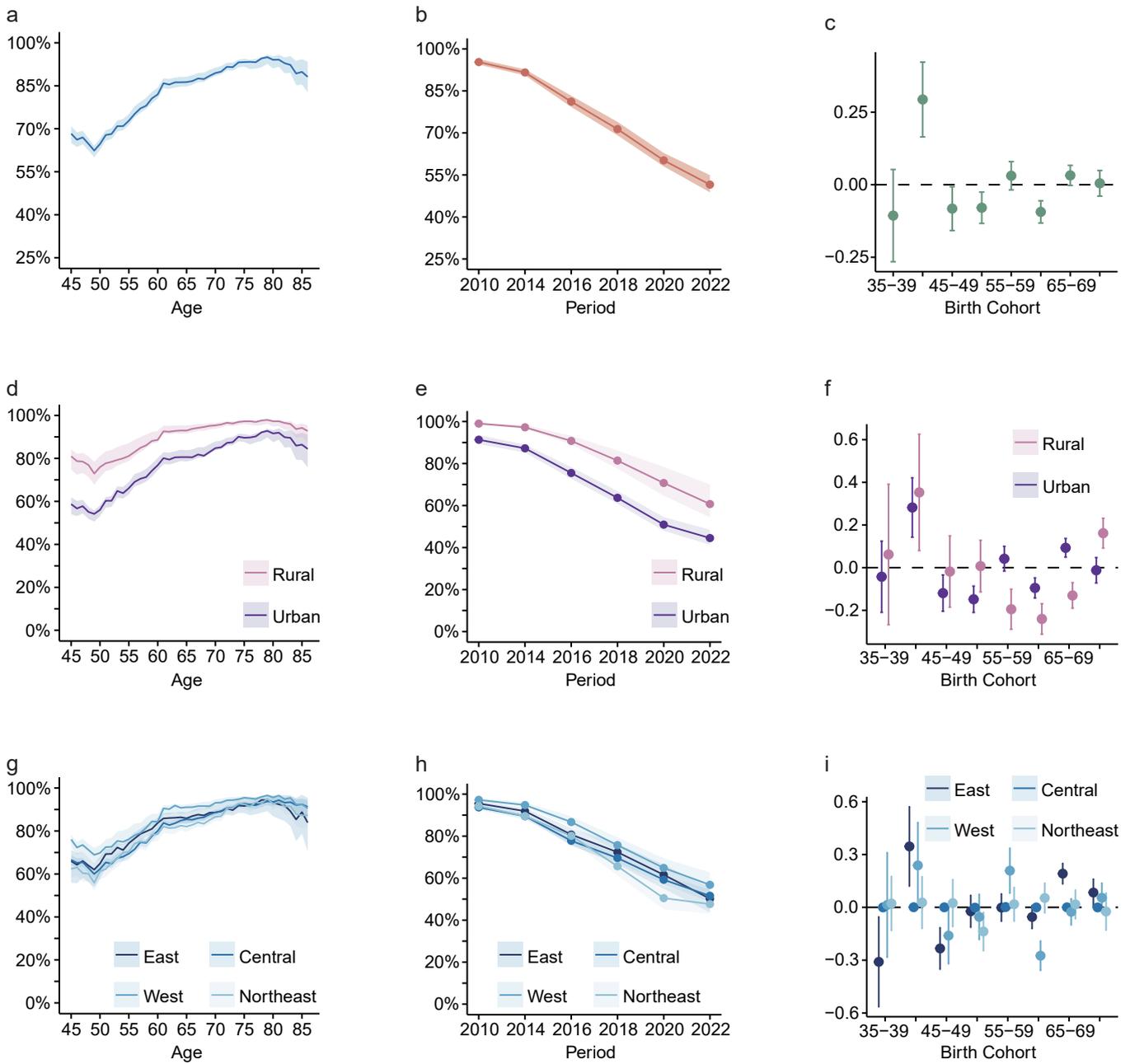

**Fig. 3 | Age–period–cohort patterns of digital exclusion in CFPS (2010–2022).**
(a) age trajectory, (b) period trend, (c) cohort deviations. (d–f) Stratified by urban–rural residence. (g–i) Stratified by region. In the age and period panels, the y-axis indicates the predicted probability of digital exclusion from weighted HAPC logistic models; shaded bands denote 95% confidence intervals. Cohort panels display the estimated random intercepts for each five-year birth cohort (on the log-odds scale). Birth cohorts are abbreviated by the last two digits (e.g., 39 refers to 1939). The reference line at 0 represents no deviation from the overall intercept, with positive values indicating higher risk of digital exclusion.

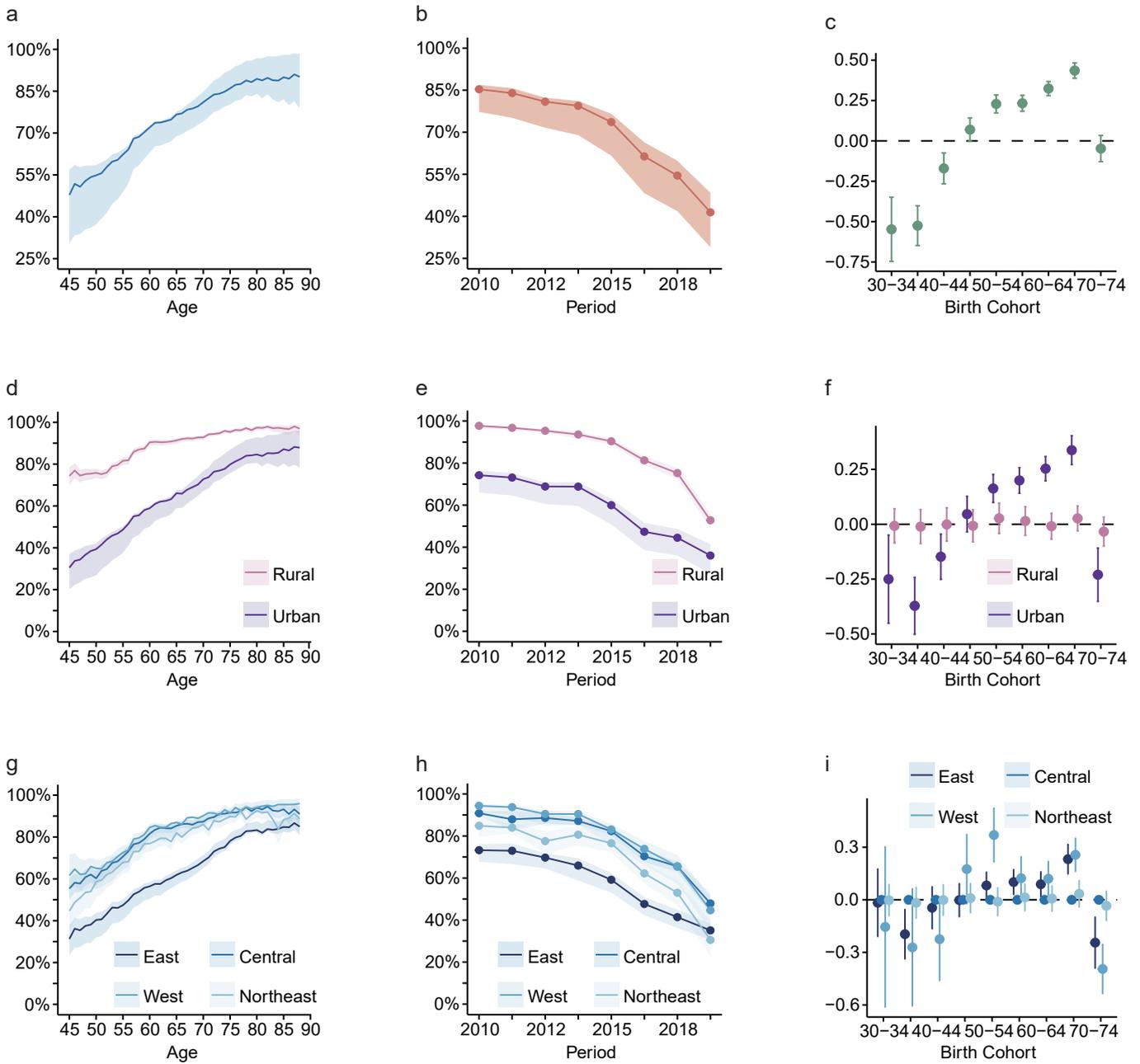

**Fig. 4 | Age–period–cohort patterns of digital exclusion in CGSS (2010–2021).**
(a) age trajectory, (b) period trend, (c) cohort deviations. (d–f) Stratified by urban–rural residence.
(g–i) Stratified by region. In the age and period panels, the y-axis indicates the predicted probability
of digital exclusion from weighted HAPC logistic models; shaded bands denote 95% confidence
intervals. Cohort panels display the estimated random intercepts for each five-year birth cohort (on
the log-odds scale). Birth cohorts are abbreviated by the last two digits (e.g., 39 refers to 1939). The
reference line at 0 represents no deviation from the overall intercept, with positive values indicating
higher risk of digital exclusion.

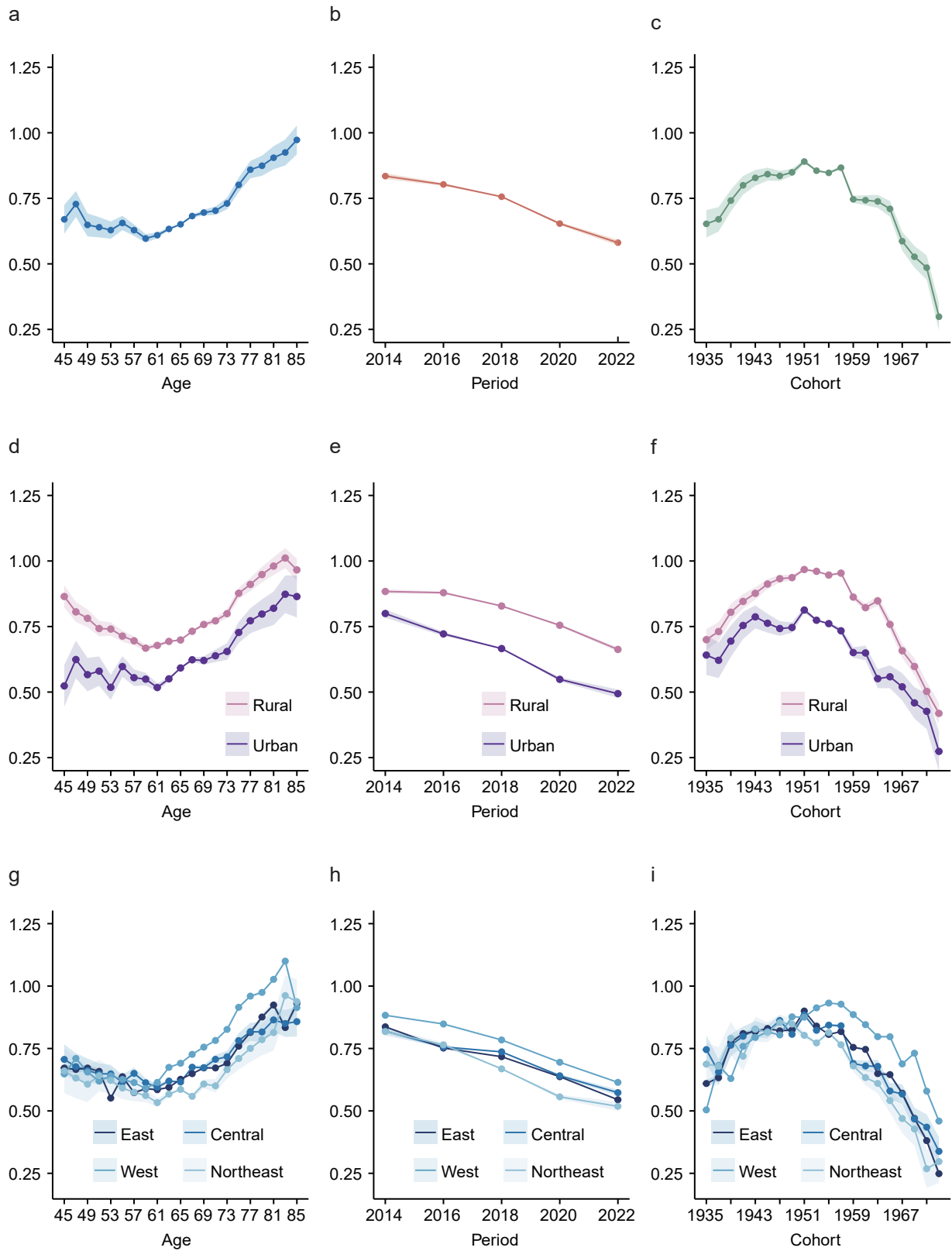

**Fig. 5 | Age–period–cohort patterns of digital exclusion from the bounding analysis in CFPS (2014–2022).**
(a) age trajectory, (b) period trend, (c) cohort deviations. (d–f) Stratified by urban–rural residence. (g–i) Stratified by region. The y-axis indicates the estimated prevalence of digital exclusion. Shaded bands represent identification bounds derived from linear constraints.

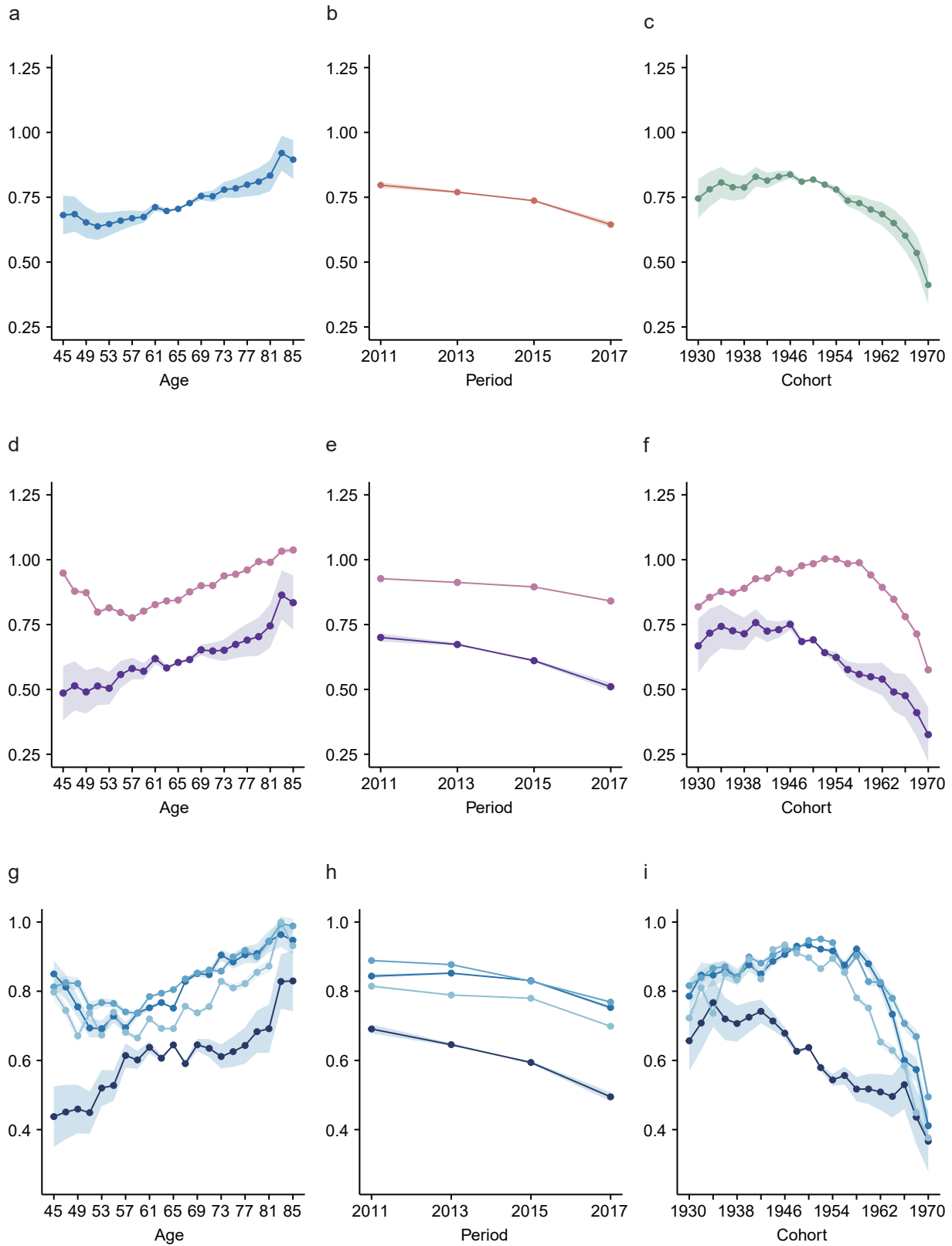

**Fig. 6 | Age–period–cohort patterns of digital exclusion from the bounding analysis in CGSS (2011–2017).**
(a) age trajectory, (b) period trend, (c) cohort deviations. (d–f) Stratified by urban–rural residence. (g–i) Stratified by region. The y-axis indicates the estimated prevalence of digital exclusion. Shaded bands represent identification bounds derived from linear constraints.

**Supplementary Materials**

**Supplementary Methods**

**Supplementary Measures**
**CHARLS**
*Hukou.* Hukou was classified into four categories agricultural hukou, non-agricultural hukou, a unified residence hukou and none hukou. We merged non-agricultural hukou and unified residence, and labeled them as urban hukou in the dataset, following the same procedure as harmonized operations.

*Education.* The education variable consists of ten categories: No formal education (illiterate), Did not finish primary school, Sishu, Elementary school, Middle school, High school, Vocational school, Two-/Three-Year College (Associate degree), Four-Year College (Bachelor's degree), and Postgraduate (Master's/PhD). In the 2020 measurement, the Postgraduate category was further divided into Master's and PhD, resulting in a total of 11 categories. Due to data sparsity in specific categories, we did not adopt the education classification used in the harmonized dataset (RAEDUCL), which groups education levels into three broad categories: less than lower secondary, upper secondary & vocational training, and tertiary. Instead, we categorized education into five groups: illiterate, lower than elementary school, elementary school, middle school, and high school or higher.

*Self-reported health.* Two variables in CHARLS measure self-reported health; we chose the one that aligns with CGSS, using a 1–5 point scale, and recoded it so that higher scores indicate better health.

*IADL.* The limitations of instrumental activities of daily living (IADLs) included managing money, taking medications, shopping for groceries, preparing meals, cleaning the house, and making phone calls. As Wave 1 did not measure difficulty in making phone calls, we included only five IADLs in our analysis. The IADL variable was coded as 0 if individuals had none of these limitations, and 1 if they had any one of them.

*Cognitive function.* In the CHARLS pilot survey and its three subsequent national surveys, the research team (Zhao et al., 2020) used a simplified version of the Telephone Interview for Cognitive Status (TICS), which is structured similarly to the Health and Retirement Study (HRS) in the United States. The key components include: Time Orientation Test: Respondents are asked to identify the current month, date, year, season (using both the lunar and Gregorian calendars), and day of the week. Self-Rated Memory: Respondents are asked to assess their memory using the following scale: Excellent, Very Good, Good, Fair, or Poor. Calculation Ability Test (Serial Subtraction): Starting from 100, respondents are required to subtract 7 in succession, up to five times. Visuospatial Ability Test: Respondents are asked to draw overlapping pentagons. Word Recall Test: This test is based on the HRS version and includes Immediate Recall and Delayed Recall. Respondents listen to a list of 10 nouns and are then asked to recall them after approximately five minutes. The delayed recall test does not involve rereading the words, in contrast to some versions of memory tests. The total score for cognitive function ranges from 1 to 36 points, with the individual components scoring as follows: Date Recognition (0-5), Self-Rated Memory (1-5), Serial Subtraction (0-5), Visuospatial Ability (0-1), Immediate Word Recall Test (0-10), and Delayed Word Recall Test (0-10), with higher scores indicating better cognitive function.

**CPFS**
Wave 2 in CFPS was not selected due to the absence of variables related to digital exclusion.

*Birth year.* The birth years in CFPS sometimes do not align with the age calculated in the respective waves, and there are instances where an individual's birth year is inconsistent across waves. To address these discrepancies, we first excluded participants (pid) who were measured in only two waves and whose birth year difference between the two waves was greater than or equal to 2 years. Next, we excluded participants whose birth year differed across multiple waves. For the remaining participants, we calculated a corrected birth year by using the most frequently occurring birth year across waves as the corrected value. In cases where there was a tie in frequency, the most recent recorded birth year was used, as it is presumed to be more accurate due to the CFPS updating birth year values in later waves. After these steps, 114 participant pids were excluded, and a thorough check confirmed that no special values remained.

*Education.* To ensure consistency in the coding of the education variable and to improve data balance, we recategorized the original CFPS education levels as follows: (1) Illiterate/functional illiterate; (2) Elementary school; (3) Middle school; (4) High school and above (including high school, technical secondary school, vocational school, junior college, undergraduate, master's, and doctoral degrees).

*ADL and IADL.* CFPS measured IADLs for respondents aged 45 years and above, and it is a modified version of the widely used Lawton IADLs scale (Lawton & Brody, 1969), by replacing communication, medication, and finance management with two basic skills (i.e., ambulating and feeding) from activities of daily living (ADLs) scale (Katz et al., 1963).

*Familysize.* In the 2014 wave of CFPS, family-size was assessed based on economic independence. Specifically, CFPS identified whether family members had formed a separate household, defined by financial autonomy. As a result, within the same pid, an individual could belong to two types of households: (1) a newly formed, economically independent household and (2) the biological family (i.e., gene family). Our measurement of family-size is based on the biological family, with the key criterion for determining household membership being economic co-dependence—conceptualized as "sharing a common stove" ("同灶吃饭"). If a family member was deemed economically independent, they were classified as part of a newly formed household. The primary criterion for determining whether individuals belonged to the same economic unit was whether they shared financial resources and meals. In subsequent CFPS waves, family-size was compiled by the CFPS team following a similar principle, where individuals were considered part of the same family if they remained economically dependent on the household unit. This corresponds to the 2014 criterion of "sharing a common stove". To maintain consistency with later waves, we removed 93 cases in Wave 3 where individuals were recorded as belonging to both a newly formed household and a biological family. Only the family-size of the biological family was retained in our dataset.

**CGSS**

*Education.* In the CGSS dataset, the coding for education is as follows: 1) No formal education; 2) Sishu or literacy class; 3) Elementary school; 4) Milddle school; 5) Vocational high school; 6) General high school; 7) Technical secondary school; 8) Junior college (adult higher education); 9) Junior college (formal higher education); 10) Bachelor's degree (adult higher education); 11) Bachelor's degree (formal higher education); 12) Graduate degree or higher; 13) Other. To ensure consistency with the other two datasets, we categorized education into four groups: illiterate and lower than elementary school, elementary school, middle school, and high school or higher.

*Subjective social class.* The subjective social class scores in the CGSS are derived from the MacArthur Scale. The MacArthur Scale is a ten-point gradient measure of subjective social class, with a score of 1 representing the lowest class and a score of 10 representing the highest class. We categorized the scores into five levels of social class identity by grouping the 1-10 range into intervals of two points, creating a classification variable with five levels.

*Social participation.* The measurement of social participation is based on the question: "In the past year,

have you frequently engaged in socializing or visiting others during your free time?" Responses indicating "never" are considered as no social activity and are coded as 0. Responses ranging from "rarely" to "very frequently" are coded as 1.

**LPA and ROC analysis**

Cognitive function (1-36 points) was measured using six components: self-reported memory (1-5points), immediate word recall (0–10 points), delayed word recall (0–10 points), serial 7s (0–5 points), orientation (0–5 points), and visuospatial ability (0–1 point). Due to the absence of a predefined cutoff for identifying cognitive impairment, the analytical strategy was conducted in two primary stages. First, Latent Profile Analysis (LPA) was applied to identify distinct cognitive subgroups within the sample. Second, Receiver Operating Characteristic (ROC) analysis was performed based on the LPA-derived subgroups to determine the optimal diagnostic cutoff score for identifying individuals at risk of cognitive impairment.

***Latent Profile Analysis*** was conducted to identify empirical subgroups of participants based on their patterns of cognitive performance, using the *tidyLPA* package in R. Participants with any missing values on the aforementioned six cognitive function indicators were excluded via listwise deletion. We estimated and compared a series of LPA models with the number of profiles ranging from one to eight. To determine the best-fitting model, we compared different LPA specifications for variances and covariances, such as constraining variances to be equal and fixing covariances to zero (Pastor et al., 2007) .The final model selection was based on statistical fit indices (lower Akaike Information Criterion [AIC] and Bayesian Information Criterion [BIC]) (Nylund-Gibson & Choi, 2018), classification quality (higher Entropy) (Ramaswamy et al., 1993), and the substantive interpretability of the profiles. Significant results from the Bootstrap Likelihood Ratio Test (BLRT) (p-value < 0.05) indicate that the model with k classes fits the data better than the model with k-1 classes (Muthén & Muthén, 2010). Each identified PIU risk subgroup should include at least 5% of the sample (Nagin, 2005; Wendt et al., 2019).The two-profile solution with equal variances and zero covariances was selected, distinguishing a "risk of cognitive impairment" group (Profile 1) and a "non-risk of cognitive impairment" group (Profile 2), based on their distinct cognitive performance. Participants were assigned to the profile with the highest posterior probability of membership.

***ROC analysis*** was then performed using the *cutpointr* package to derive an optimal cutoff point for the cognitive score, which differentiated between the risk and non-risk groups. The LPA-derived classification was used as the reference, with the "risk of cognitive impairment" profile designated as the positive class. The optimal cutoff was defined as the value that maximized Youden's index (Akobeng, 2007). To ensure robustness and stability, a bootstrapping procedure with 1,000 resamples was implemented, and the cutoff optimization was based on the mean metric across all bootstrap iterations. The threshold was defined as a score less than or equal to the cutoff value, classifying individuals into the risk group. The classification performance was evaluated using sensitivity, specificity, accuracy, area under the curve (AUC), and Youden's index (Carter et al., 2016). All analyses were conducted on a MacBook with Apple M2 Pro chip using R version 4.5.0.

**Results**

The LPA results supported a two-profile model, identifying a risk group and a low-risk group (see figure SM 1 and table SM1). Figure SM 2 shows the optimal cutpoint and ROC curve for cognitive risk classification. The ROC analysis yielded a highly stable optimal cutoff point of 18. This threshold demonstrated excellent discriminatory power, with a median out-of-bag AUC of 0.99. At this cutoff, the median bootstrapped sensitivity was 0.97, indicating that the test correctly identified 97% of individuals in the risk group. The median bootstrapped specificity was 0.92, meaning it correctly identified 92% of individuals in the non-risk group. The corresponding median bootstrapped Youden's index was 0.89, and the out-of-bag accuracy was 0.94 (see Table SM2). Finally, the sample size for those without cognitive risk is 34,923, and for those with cognitive risk, it is 30,145.

**Table SM1. Fit indices and LPA results with Model 1 and its class solutions indicating AIC, BIC,**

**Entropy, and BIRT.**

| Model | Class | AIC | BIC | Entropy | Min.probability | Smallest (%) | n | BLRT (*p*) |
|-------|-------|-----|-----|---------|-----------------|--------------|---|-----------|
| 1 | 1 | 1268769.38 | 1268878.38 | 1.00 | 1.00 | 1.00 | | |
| | 2 | 1232958.95 | 1233131.53 | 0.66 | 0.87 | 0.43 | | 0.01 |

*Note.* Model 1 with equal variances and zero covariances was selected. Min. probability: the minimal probability value for assignment to one of the classes within this specific class solution.

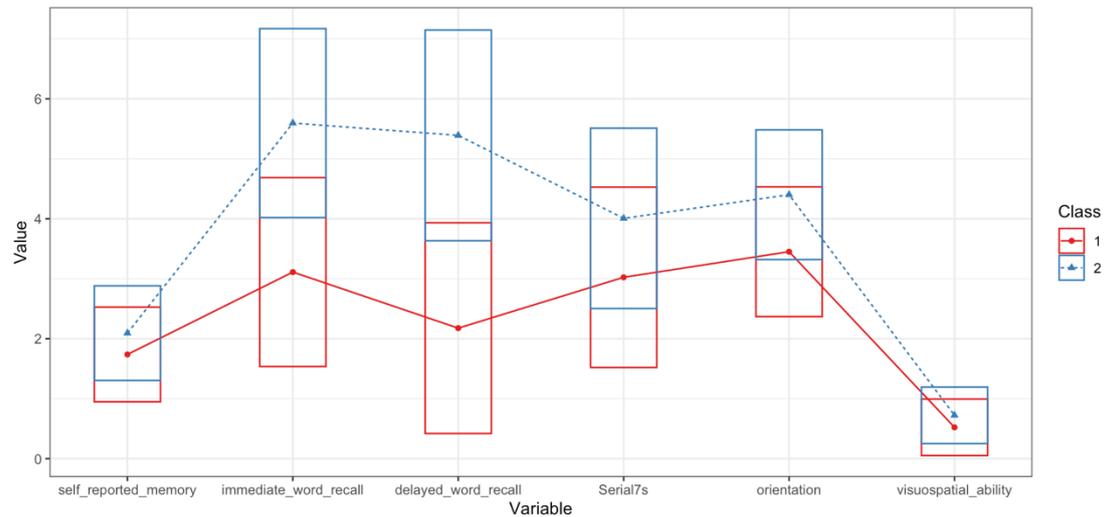

**Figure SM 1. Latent profile analysis (LPA) results based on six cognitive indicators.** Two distinct latent classes were identified. Class 1 (red solid line) represents a cognitively high-risk group characterized by lower performance across all cognitive domains. Class 2 (blue dashed line) reflects a cognitively low-risk group with relatively higher cognitive functioning. The boxes indicate the variability (standard deviation) around the mean values for each cognitive task.

**Table SM2. Bootstrap summary of optimal cutpoint and diagnostic accuracy metrics across 1,000 iterations.**

| Variable | Min. | 1st Qu. | Median | Mean | 3rd Qu. | 95% | Max. | SD |
|----------|------|---------|--------|------|---------|-----|------|-----|
| Optimal_cutpoint | 18.00 | 18.00 | 18.00 | 18.00 | 18.00 | 18.00 | 18.00 | 0 |
| AUC_b | 0.99 | 0.99 | 0.99 | 0.99 | 0.99 | 0.99 | 0.99 | 0 |
| AUC_oob | 0.99 | 0.99 | 0.99 | 0.99 | 0.99 | 0.99 | 0.99 | 0 |
| youden_b | 0.88 | 0.89 | 0.89 | 0.89 | 0.89 | 0.89 | 0.90 | 0 |
| youden_oob | 0.88 | 0.89 | 0.89 | 0.89 | 0.89 | 0.89 | 0.90 | 0 |
| acc_b | 0.94 | 0.94 | 0.94 | 0.94 | 0.94 | 0.94 | 0.94 | 0 |
| acc_oob | 0.94 | 0.94 | 0.94 | 0.94 | 0.94 | 0.94 | 0.95 | 0 |
| sensitivity_b | 0.97 | 0.97 | 0.97 | 0.97 | 0.97 | 0.97 | 0.98 | 0 |
| sensitivity_oob | 0.97 | 0.97 | 0.97 | 0.97 | 0.97 | 0.97 | 0.98 | 0 |
| specificity_b | 0.91 | 0.92 | 0.92 | 0.92 | 0.92 | 0.92 | 0.92 | 0 |
| specificity_oob | 0.91 | 0.91 | 0.92 | 0.92 | 0.92 | 0.92 | 0.92 | 0 |

Note: _b = within-bootstrap estimate; _oob = out-of-bag estimate. Estimates were identical across resamples.

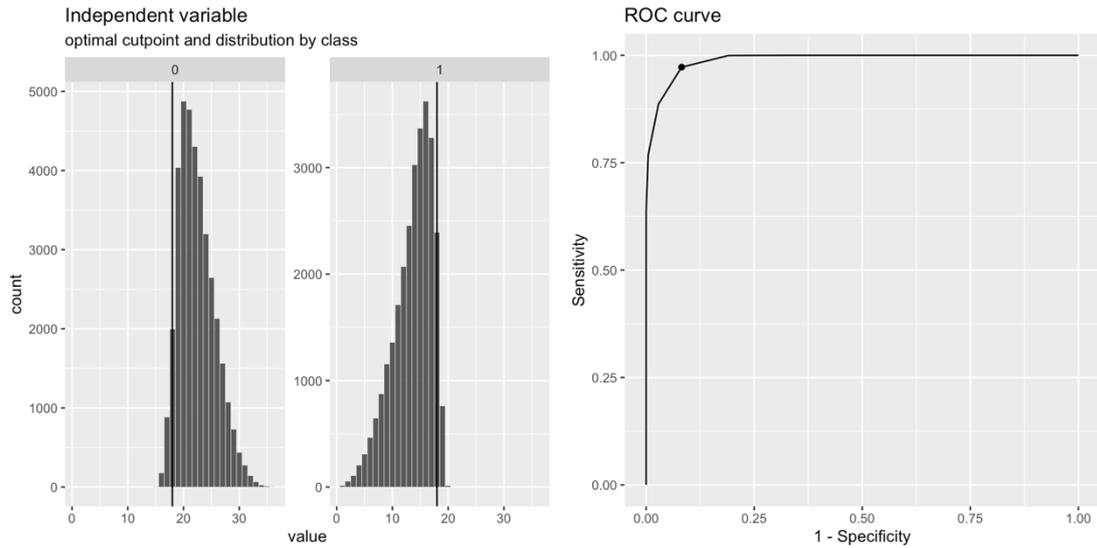

**Figure SM2. Optimal Cutpoint and ROC Curve for Cognitive Risk Classification.** The left panel shows the distribution of the continuous cognitive score by class (0 = low-risk group, 1 = high-risk group), along with the optimal cutpoint determined by maximizing the sum of sensitivity and specificity. The right panel displays the ROC curve for this classification, illustrating the trade-off between sensitivity and 1-specificity across all possible thresholds. The curve indicates high discriminative performance for the selected cognitive composite.

**Covariate balance assessment**

    We evaluated whether the weighting strategy improved covariate balance between respondents with observed digital-exclusion items and those with item nonresponse, using the *cobalt* package (Greifer, 2020). Balance was assessed using standardized mean differences (SMDs) with pooled standard deviations, comparing survey weights alone versus the combined weights (wave-specific cross-sectional survey weight multiplied by stabilized IPW for digital-exclusion item response). We considered an absolute SMD below 0.25 to indicate acceptable balance and additionally report balance using 0.10 as a more stringent benchmark. Figures SM3–SM7 present the SMDs before and after applying the weights for each of the three datasets.

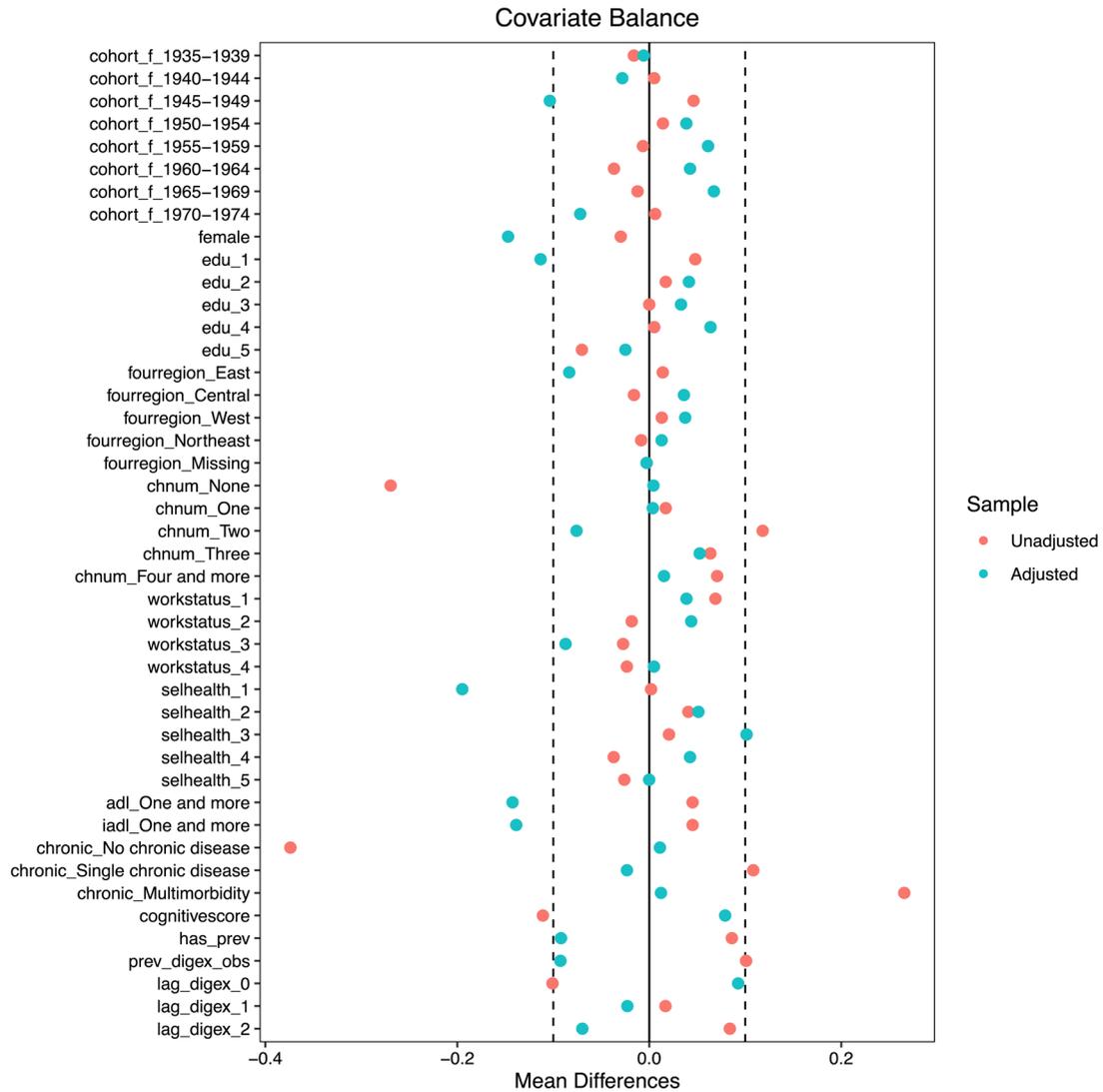

**Figure SM3.** Covariate Balance After Weighting for Digital Exclusion Response (CHARLS)

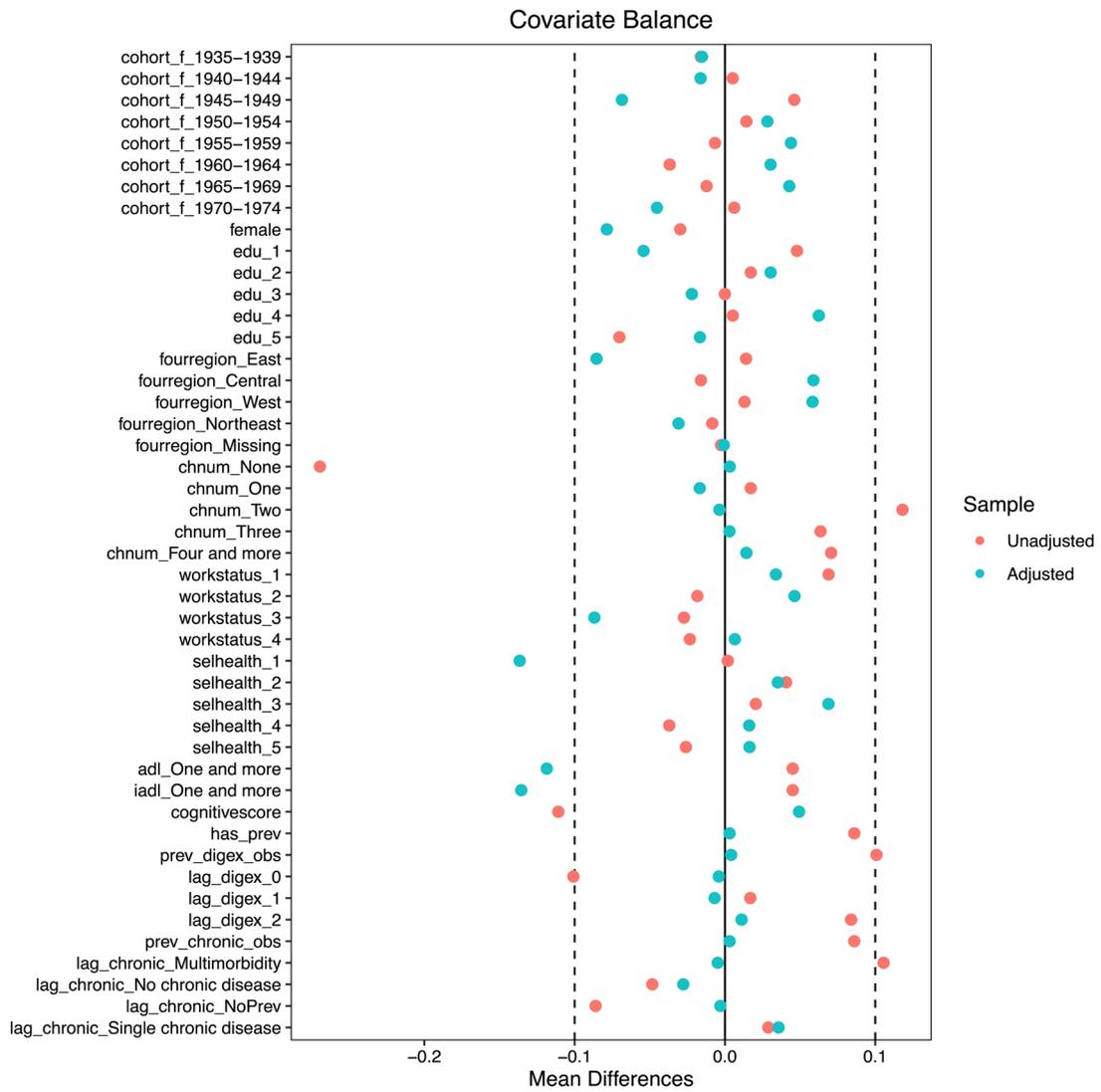

**Figure SM4.** Covariate Balance After Weighting for Digital Exclusion and Multimorbidity Status Response (CHARLS)

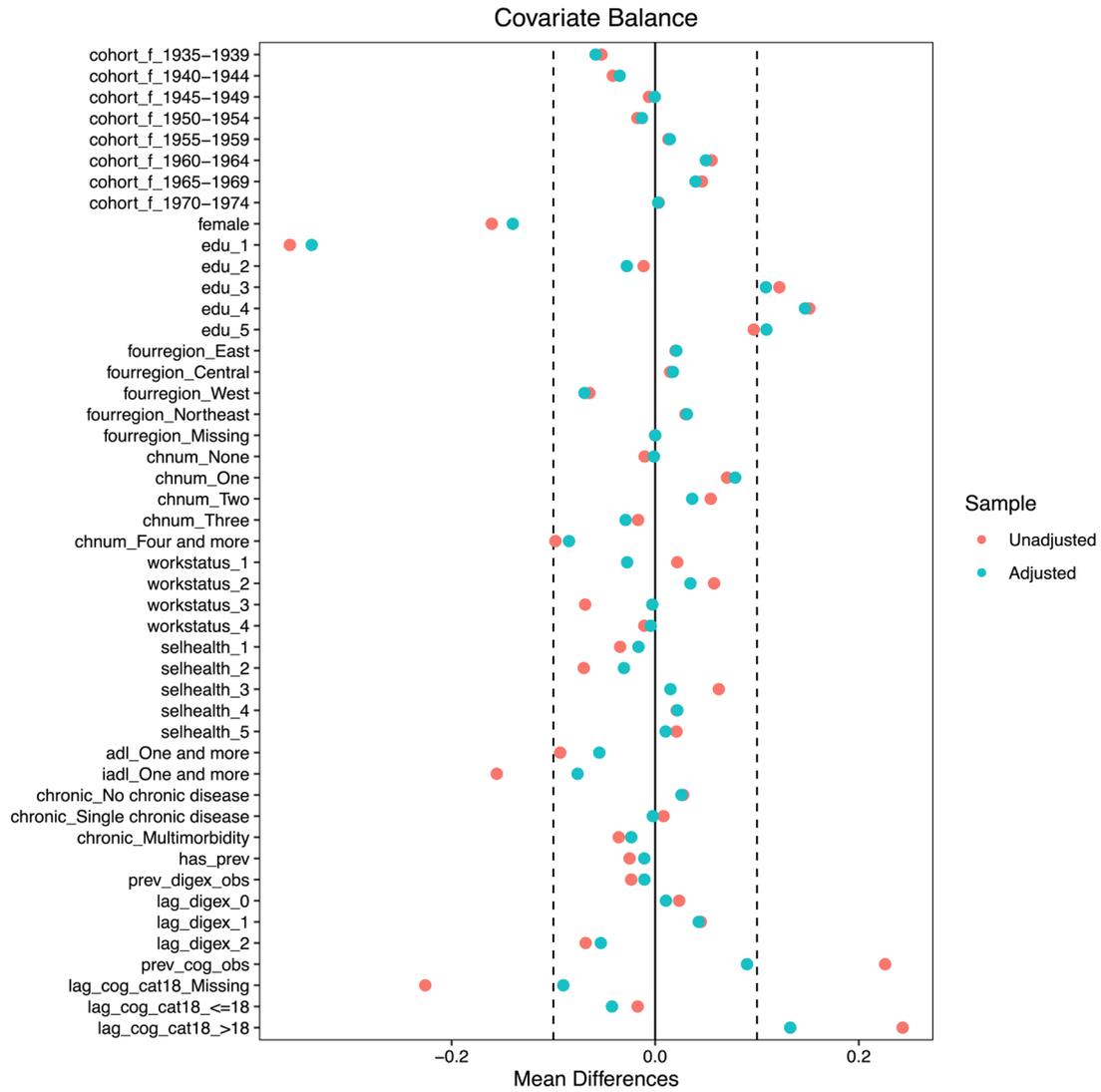

**Figure SM5.** Covariate Balance After Weighting for Digital Exclusion and Cognitive Function Response (CHARLS)

*Note.* Although a few education indicators exceeded |SMD| = 0.25, we retained the combined weights because they substantially reduced imbalance across most covariates and were constructed specifically to correct joint item nonresponse in digital exclusion and cognitive function, for which education is a primary determinant. Lower educational attainment appears to be associated with item nonresponse, which can limit covariate overlap and make education harder to fully balance.

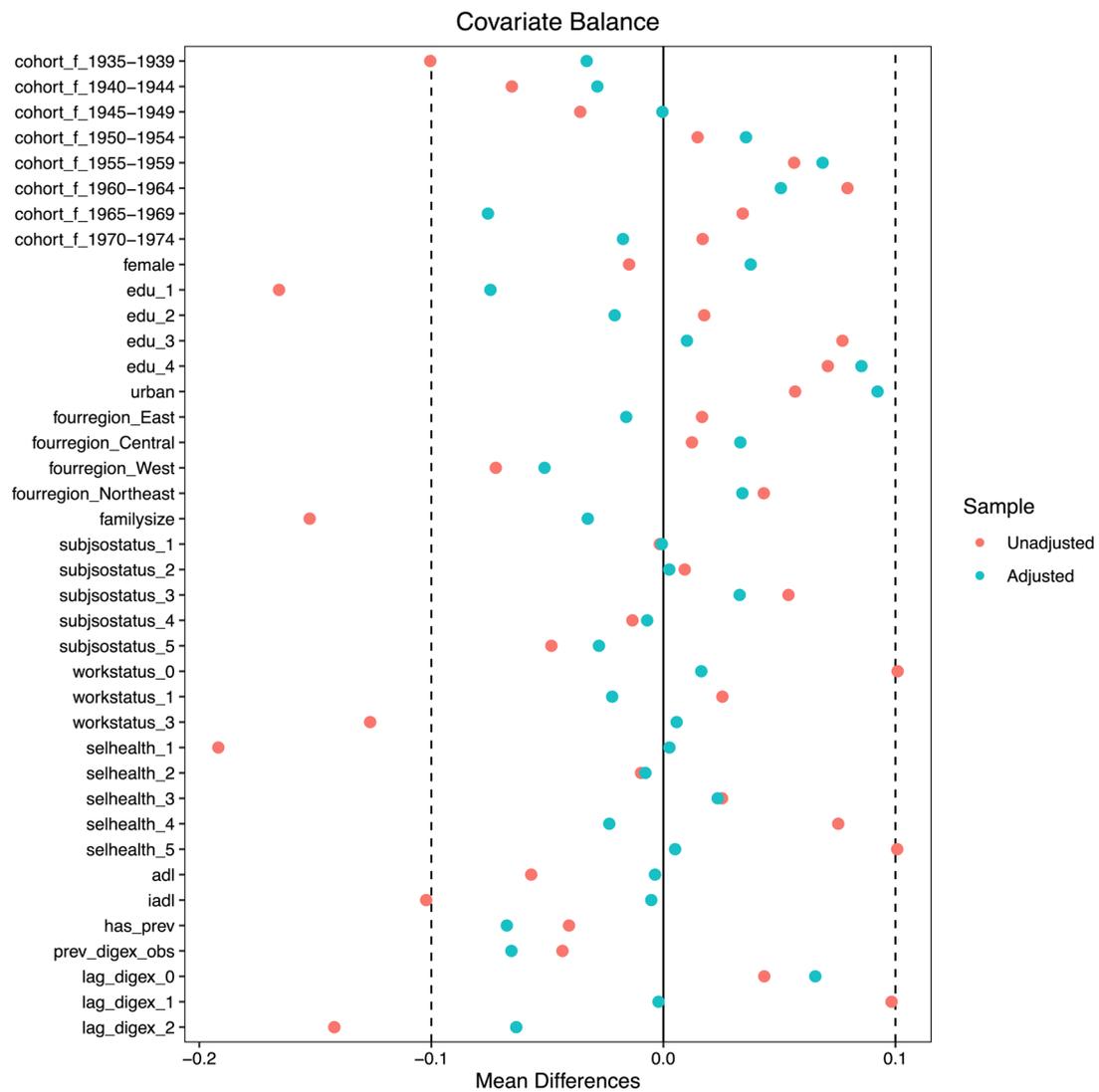

**Figure SM6.** Covariate Balance After Weighting for Digital Exclusion Response (CFPS)

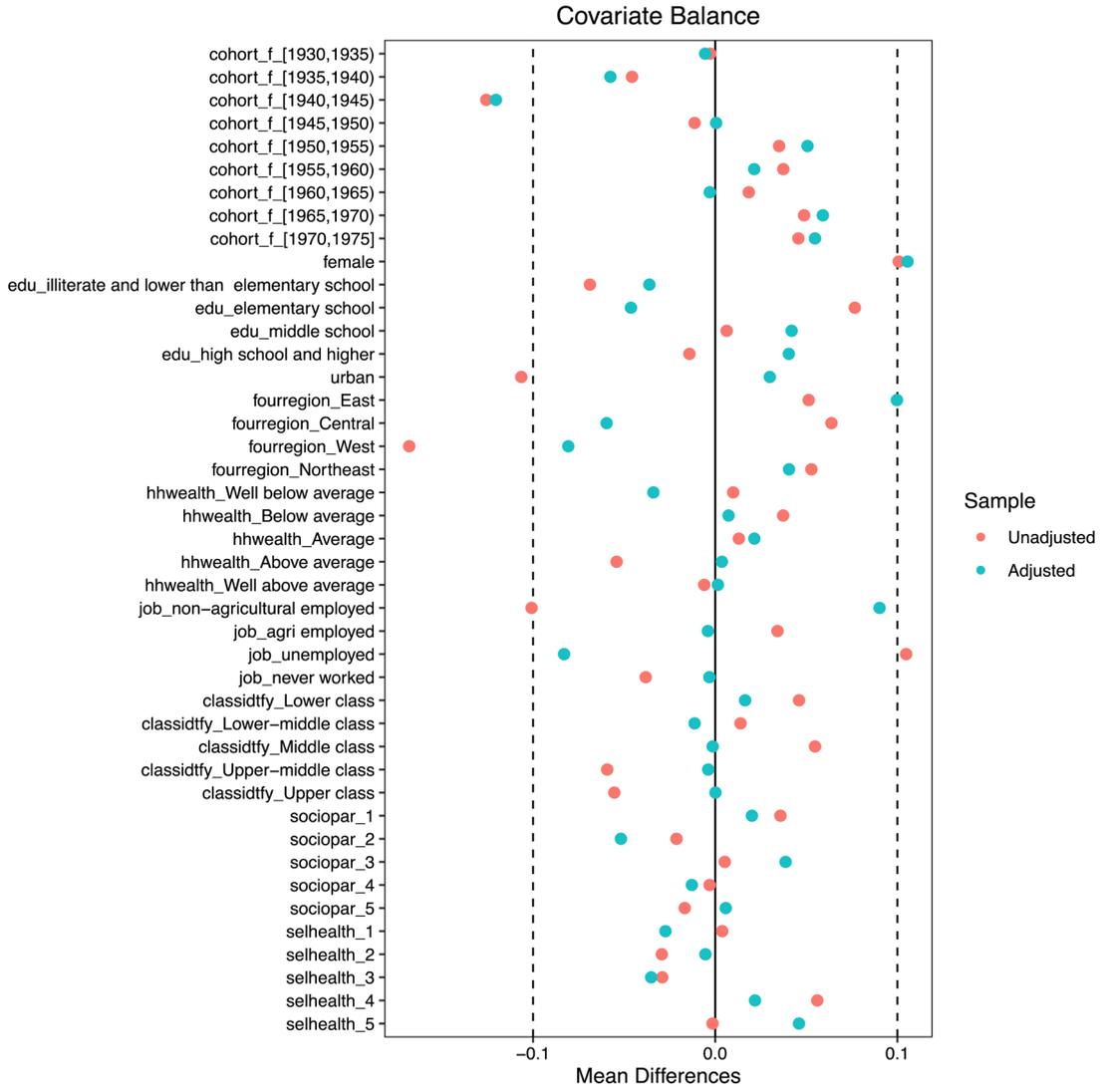

**Figure SM7.** Covariate Balance After Weighting for Digital Exclusion Response (CGSS)

**Canonical solution line**

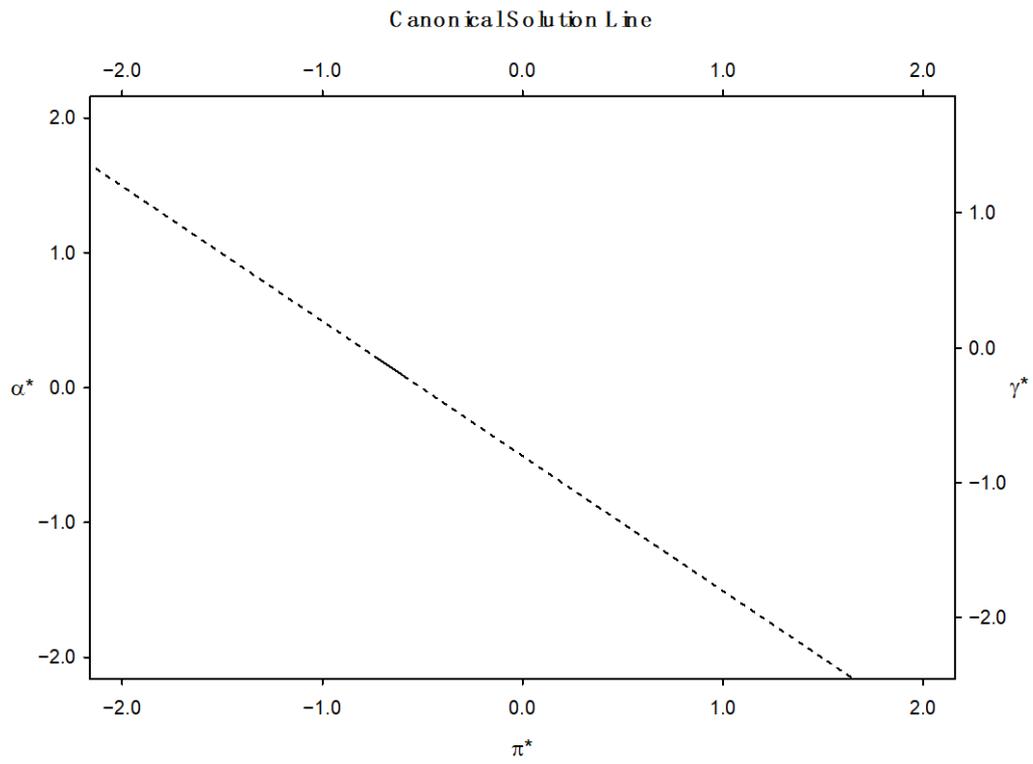

**Figure S1. Canonical solution line, CFPS 2014–2022**

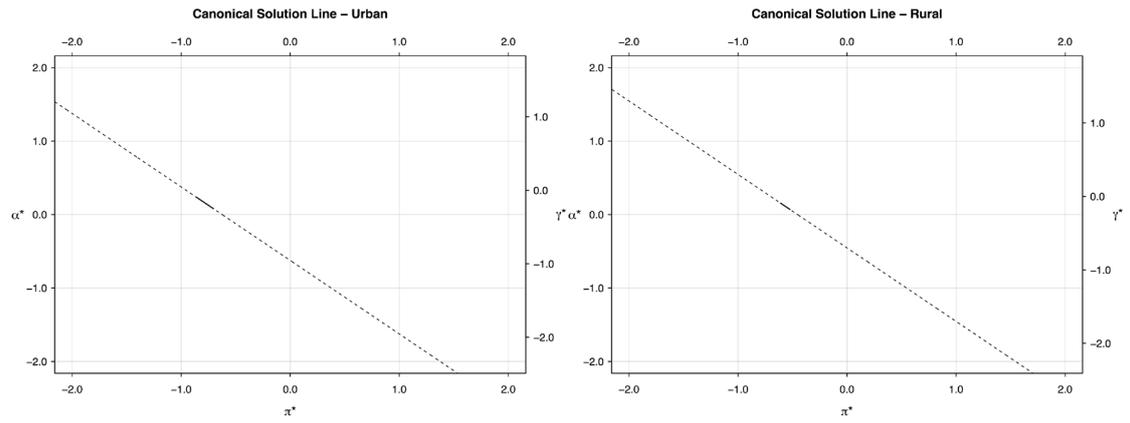

**Figure S2. Canonical solution line by residence, CFPS 2014–2022.**

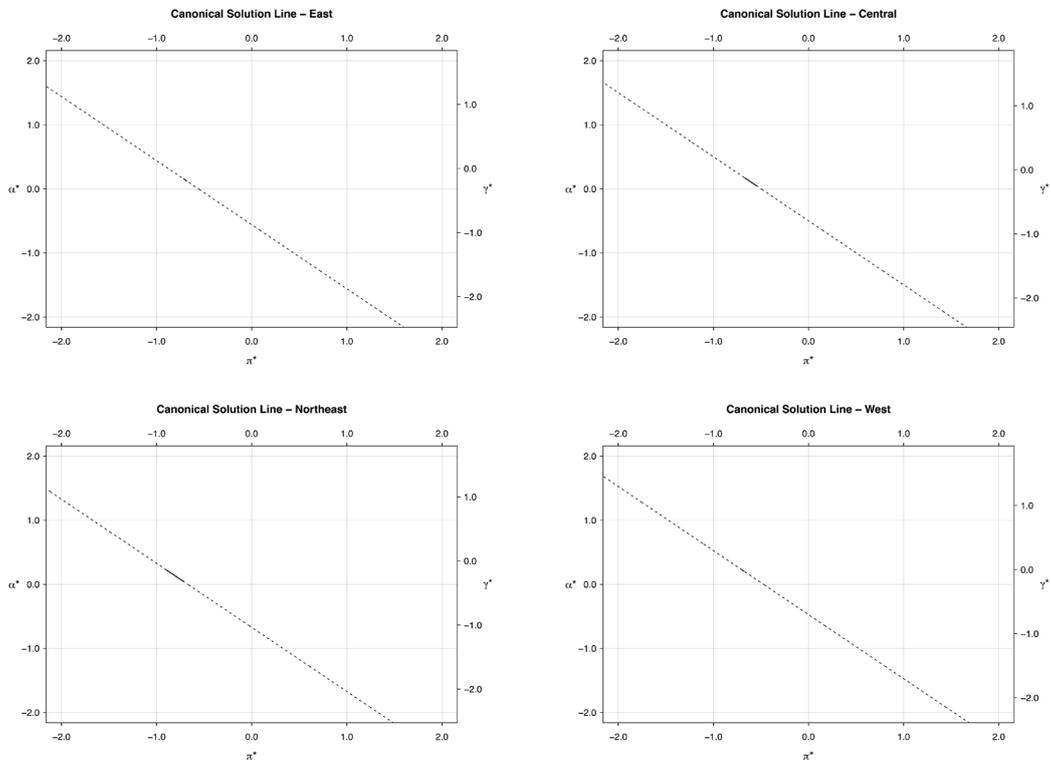

**Figure S3. Canonical solution line by region, CFPS 2014–2022.**

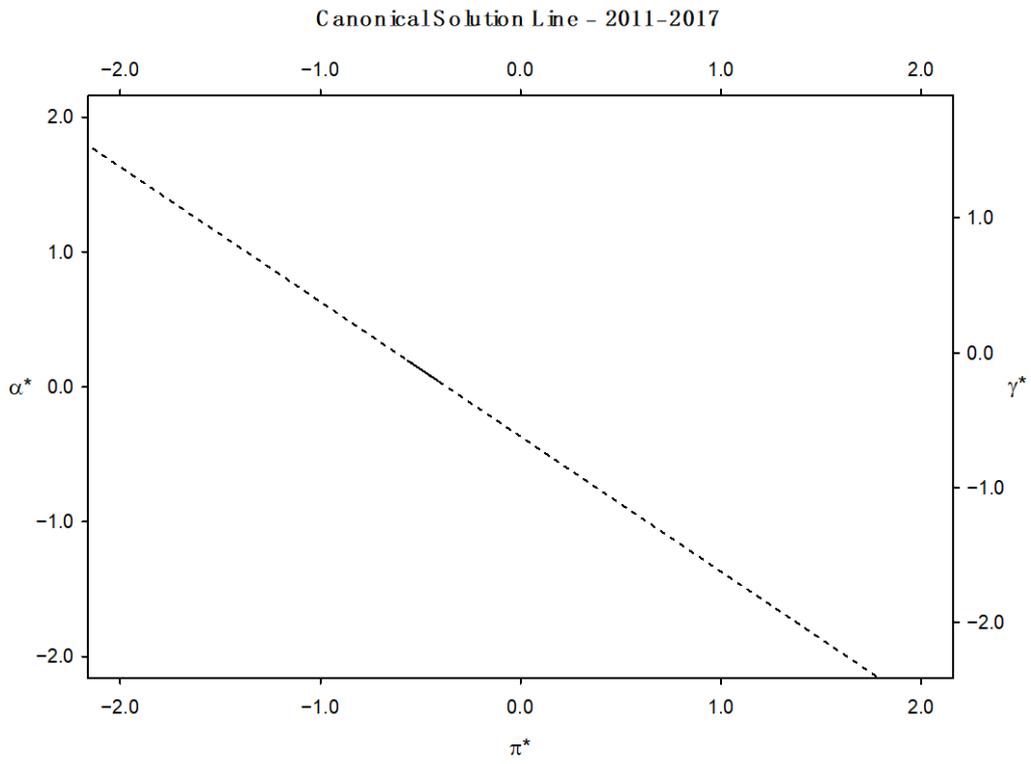

**Figure S4. Canonical solution line, CGSS 2011–2017.**

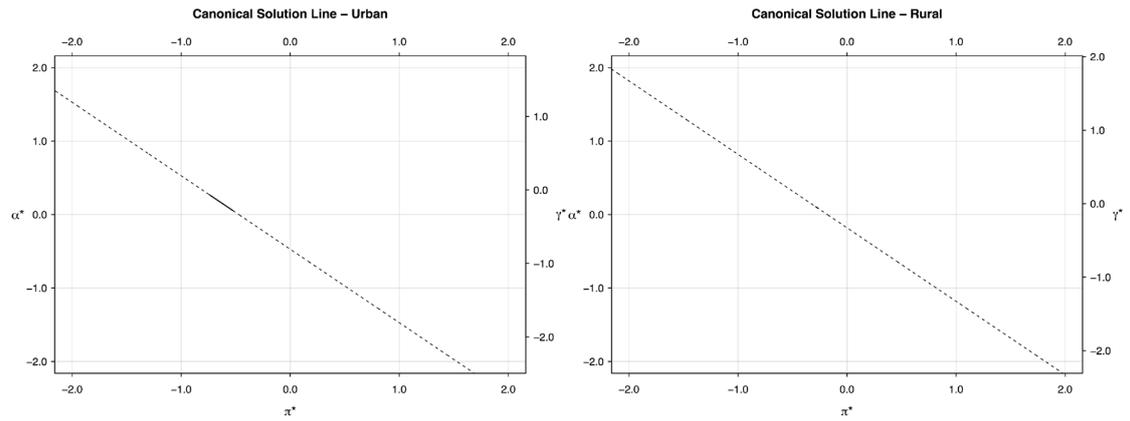

**Figure S5. Canonical solution line by residence, CGSS 2011–2017.**

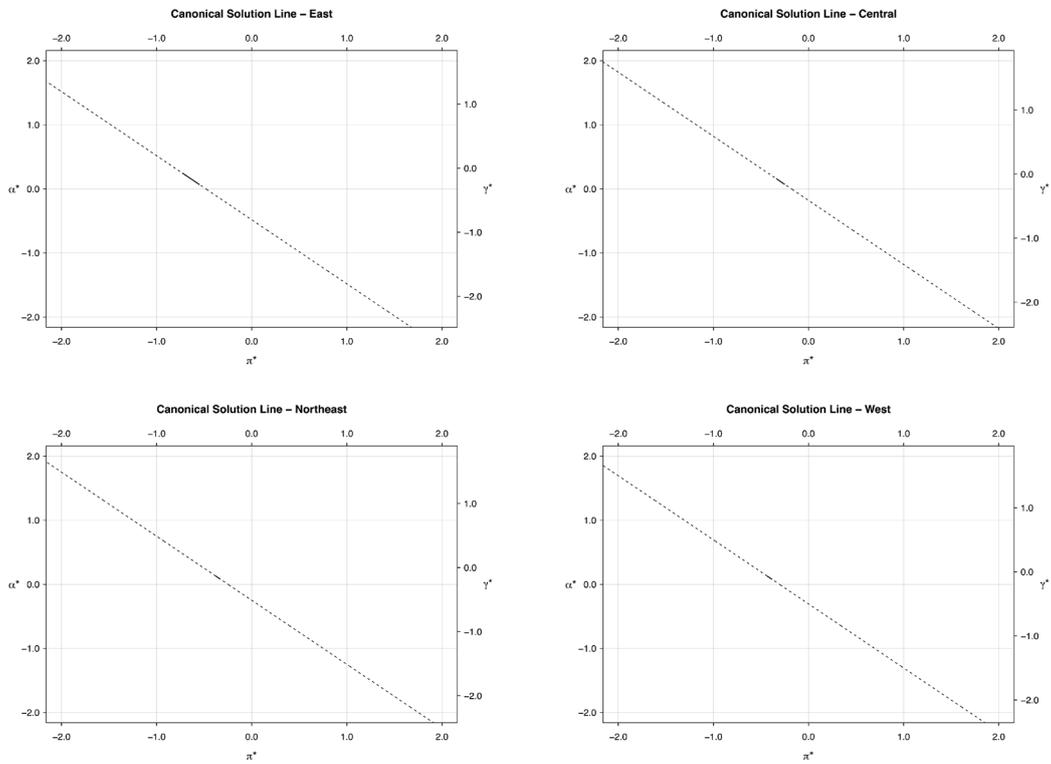

**Figure S6. Canonical solution line by region, CGSS 2011–2017.**

**Table S1 Harmonization of Digital Exclusion Across Datasets**

| Dataset | Measure | Options | Binary recode (0/1) |
|---|---|---|---|
| CHARLS | Any Internet use in the past month | Yes; No | 1 = digitally excluded if No; 0 = not excluded if Yes; |
| CFPS | 2010 & 2014: Any Internet use. 2016/2018/2022: Internet use by device (mobile / computer) | Yes; No | 2010/2014: 1 if No, 0 if Yes. 2016/2018/2022: 1 if neither mobile nor computer; 0 if mobile or computer (either); |
| CGSS | Internet use frequency in the past year (incl. mobile Internet) | Never; Rarely; Sometimes; Often; Very often | 1 = digitally excluded if Never; 0 = not excluded if Rarely/Sometimes/Often/Very often |

**Table S2 Overview of analysis weights across models**

| Model | Dataset (s) | Analytic sample | Analysis weight |
|---|---|---|---|
| Weighted HAPC models | CHARLS, CFPS, CGSS | Interview-complete respondents with available wave-specific cross-sectional survey weights | Wave-specific cross-sectional survey weight multiplied by stabilized IPW for digital-exclusion item response |
| Weighted HAPC models stratified by multimorbidity status | CHARLS | Interview-complete respondents with available wave-specific cross-sectional survey weights and joint availability of digital exclusion and multimorbidity status | Wave-specific cross-sectional survey weight multiplied by stabilized IPW for digital-exclusion and multimorbidity-status response |
| Weighted HAPC models stratified by cognitive function | CHARLS | Interview-complete respondents with available wave-specific cross-sectional survey weights and joint availability of digital exclusion and cognitive function | Wave-specific cross-sectional survey weight multiplied by stabilized IPW for digital-exclusion and cognitive-function response |
| Bounding analysis | CHARLS, CFPS, CGSS | Aggregated age–period–cohort cells derived from the primary analytic sample (interview-complete respondents with wave-specific cross-sectional survey weights) | Within each cell, digital exclusion was calculated as a weighted mean using the final standardized combined weight (wave-specific survey weight × stabilized IPW for digital-exclusion item response) |

**Table S3 Unweighted Descriptive Statistics of the Study Samples**

|  | CHARLS (N = 81,451) | CFPS (N = 85,469) | CGSS (N = 50,721) |
|---|---|---|---|
| Digital Exclusion | 70831 (86.96) | 67185 (78.61) | 35100 (69.20) |
| Age (years) | 59.77 (8.89) | 58.57 (9.02) | 60.32 (10.15) |
| Cohort |  |  |  |
| 1930-1934 |  |  | 1811 (3.57) |
| 1935-1939 | 4358 (5.35) | 3221 (3.77) | 3049 (6.01) |
| 1940-1944 | 7161 (8.79) | 5545 (6.49) | 4356 (8.59) |
| 1945-1949 | 11040 (13.55) | 9565 (11.19) | 6255 (12.33) |
| 1950-1954 | 15724 (19.30) | 14023 (16.41) | 8608 (16.97) |
| 1955-1959 | 14293 (17.55) | 13763 (16.10) | 8236 (16.24) |
| 1960-1964 | 15347 (18.84) | 15866 (18.56) | 8739 (17.23) |
| 1965-1969 | 11679 (14.34) | 15320 (17.92) | 6863 (13.53) |
| 1970-1974 | 1849 (2.27) | 8166 (9.55) | 2804 (5.53) |
| Sex |  |  |  |
| Male | 39101 (48.01) | 42588 (49.83) | 24642 (48.58) |
| Female | 42350 (51.99) | 42881 (50.17) | 26079 (51.42) |
| Residence |  |  |  |
| Rural | 62949 (77.28) | 45147 (52.82) | 21006 (41.41) |
| Urban | 18502 (22.72) | 40322 (47.18) | 29715 (58.59) |
| Region |  |  |  |
| East | 25686 (31.54) | 28127 (32.91) | 18698 (36.86) |
| Central | 23243 (28.54) | 20897 (24.45) | 13479 (26.57) |
| West | 26681 (32.76) | 23254 (27.21) | 12451 (24.55) |
| Northeast | 5785 (7.10) | 13191 (15.43) | 6093 (12.01) |
| Missing | 56 (0.07) |  |  |

*Note.* Data are presented as Mean (standard deviation) or Frequency (%). Statistics are unweighted.

**Table S4 Unweighted Multilevel Logistic Regression of Digital Exclusion in CHARLS**

| | Full sample | Urban | Rural | East | Central | West | Northeast |
|---|---|---|---|---|---|---|---|
| Intercept | 12.331 [8.700, 17.476]*** | 2.614 [2.314, 2.954]*** | 50.147 [41.193, 61.047]*** | 8.443 [6.411, 11.120]*** | 10.267 [8.214, 12.833]*** | 15.577 [13.229, 18.341]*** | 8.388 [5.868, 11.989]*** |
| Age | 1.042 [1.008,1.077]* | 1.085 [1.073,1.098]*** | 1.147 [1.101,1.195]*** | 1.048 [0.991,1.109] | 1.069 [1.004,1.138]* | 1.119 [1.089,1.150]*** | 1.095 [1.047,1.144]*** |
| Age$^2$ | 1.002 [1.001,1.002]*** | 1.000 [1.000,1.001] | 1.000 [0.999,1.002] | 1.002 [1.000,1.003]* | 1.002 [1.000,1.003] | 1.000 [0.999,1.001] | 1.000 [0.999,1.002] |
| Period (ref: 2011) | | | | | | | |
| 2013 | 0.641 [0.556,0.740]*** | 0.685 [0.614,0.764]*** | 0.339 [0.266,0.432]*** | 0.688 [0.564,0.840]*** | 0.566 [0.445,0.720]*** | 0.588 [0.462,0.747]*** | 0.601 [0.416,0.869]** |
| 2015 | 0.393 [0.335,0.462]*** | 0.424 [0.368,0.490]*** | 0.128 [0.105,0.157]*** | 0.444 [0.362,0.544]*** | 0.339 [0.284,0.405]*** | 0.309 [0.236,0.403]*** | 0.312 [0.231,0.420]*** |
| 2018 | 0.168 [0.131,0.215]*** | 0.218 [0.181,0.263]*** | 0.030 [0.021,0.043]*** | 0.190 [0.131,0.274]*** | 0.137 [0.101,0.186]*** | 0.106 [0.087,0.128]*** | 0.131 [0.086,0.199]*** |
| 2020 | 0.028 [0.021,0.038]*** | 0.052 [0.043,0.064]*** | 0.003 [0.002,0.005]*** | 0.021 [0.024,0.058]*** | 0.021 [0.016,0.028]*** | 0.015 [0.013,0.016]*** | 0.021 [0.015,0.029]*** |
| Sex (Female = 1) | 1.617 [1.563,1.673]*** | 1.323 [1.214,1.442]*** | 1.837 [1.767,1.910]*** | 1.646 [1.535,1.765]*** | 1.698 [1.503,1.919]*** | 1.684 [1.550,1.830]*** | 1.173 [0.988,1.392] |
| *Random effects* | | | | | | | |
| σ$^2$ Cohort | 0.065 [0.017, 0.252] | 0.000 [0.000, 0.020] | 0.018 [0.005, 0.070] | 0.092 [0.019, 0.450] | 0.015 [0.002, 0.095] | 0.000 [0.000, 0.000] | 0.013 [0.005, 0.035] |
| BIC | 46271.1 | 17041.1 | 23927.4 | 15690.8 | 13592.0 | 12725.1 | 3980.0 |
| AIC | 46206.0 | 16978.5 | 23864.0 | 15625.6 | 13527.5 | 12667.8 | 3926.7 |
| N | 81,451 | 18,502 | 62,949 | 25,686 | 23,243 | 26,681 | 5,785 |

*Note.* Age was centered at 45. Fixed-effect estimates are reported as odds ratios (ORs) with 95% confidence intervals. Models are unweighted, and standard errors are clustered at the cohort level. AIC = Akaike information criterion; BIC = Bayesian information criterion. *** p < .001, ** p < .01, * p < .05.

**Table S5 Unweighted Multilevel Logistic Regression of Digital Exclusion in CFPS**

| | Full sample | Urban | Rural | East | Central | West | Northeast |
|---|---|---|---|---|---|---|---|
| Intercept | 6.995 [5.735, 8.409]*** | 3.404 [2.826, 4.101]*** | 35.201 [24.234, 51.131]*** | 4.454 [3.434, 5.775]*** | 6.984 [5.866, 8.316]*** | 16.052 [13.418, 19.203]*** | 6.717 [5.260, 8.578]*** |
| Age | 1.120 [1.095,1.146]*** | 1.114 [1.088,1.140]*** | 1.201 [1.172,1.230]*** | 1.115 [1.068,1.165]*** | 1.121 [1.102,1.141]*** | 1.137 [1.114,1.161]*** | 1.141 [1.116,1.167]*** |
| Age$^2$ | 0.999 [0.999,1.000] | 0.999 [0.999,1.000] | 0.998 [0.997,0.999]*** | 0.999 [0.998,1.001] | 1.000 [0.999,1.000] | 0.999 [0.998,1.000]* | 0.999 [0.998,1.000]** |
| Period (ref: 2010) | | | | | | | |
| 2014 | 0.494 [0.432,0.564]*** | 0.514 [0.451,0.587]*** | 0.280 [0.250,0.313]*** | 0.542 [0.479,0.613]*** | 0.460 [0.406,0.521]*** | 0.390 [0.313,0.487]*** | 0.480 [0.392,0.587]*** |
| 2016 | 0.199 [0.163,0.244]*** | 0.233 [0.195,0.278]*** | 0.063 [0.054,0.074]*** | 0.227 [0.176,0.292]*** | 0.173 [0.149,0.201]*** | 0.136 [0.111,0.166]*** | 0.195 [0.164,0.232]*** |
| 2018 | 0.090 [0.072,0.113]*** | 0.119 [0.099,0.144]*** | 0.020 [0.017,0.024]*** | 0.114 [0.084,0.154]*** | 0.086 [0.074,0.100]*** | 0.046 [0.038,0.056]*** | 0.083 [0.072,0.096]*** |
| 2020 | 0.046 [0.036,0.059]*** | 0.064 [0.051,0.080]*** | 0.009 [0.007,0.011]*** | 0.063 [0.045,0.088]*** | 0.044 [0.037,0.052]*** | 0.024 [0.019,0.030]*** | 0.033 [0.026,0.042]*** |
| 2022 | 0.028 [0.022,0.035]*** | 0.040 [0.032,0.050]*** | 0.005 [0.003,0.007]*** | 0.034 [0.024,0.048]*** | 0.027 [0.023,0.032]*** | 0.015 [0.011,0.020]*** | 0.023 [0.020,0.027]*** |
| Sex (Female = 1) | 1.447 [1.322,1.585]*** | 1.406 [1.239,1.595]*** | 1.684 [1.532,1.852]*** | 1.657 [1.459,1.883]*** | 1.501 [1.374,1.639]*** | 1.618 [1.417,1.847]*** | 0.964 [0.783,1.187] |
| *Random effects* | | | | | | | |
| σ$^2$ Cohort | 0.007 [0.001, 0.036] | 0.011 [0.005, 0.024]* | 0.048 [0.006, 0.406] | 0.015 [0.004, 0.057] | 0.000 [0.000, 0.000] | 0.022 [0.008, 0.064] | 0.004 [0.001, 0.013] |
| BIC | 69709.2 | 39394.3 | 26029.6 | 24163.8 | 17104.8 | 16441.2 | 11221.5 |
| AIC | 69634.3 | 39334.0 | 25959.9 | 24106.1 | 17041.2 | 16384.8 | 11161.6 |

| N | 85,469 | 40,322 | 45,147 | 28,127 | 20,897 | 23,254 | 13,191 |

*Note.* Age was centered at 45. Fixed-effect estimates are reported as odds ratios (ORs) with 95% confidence intervals. Models are unweighted, and standard errors are clustered at the cohort level. AIC = Akaike information criterion; BIC = Bayesian information criterion. *** p < .001, ** p < .01, * p < .05.

**Table S6 Unweighted Multilevel Logistic Regression of Digital Exclusion in CGSS**

| | Full sample | Urban | Rural | East | Central | West | Northeast |
|---|---|---|---|---|---|---|---|
| Intercept | 1.383 [1.006, 1.901]* | 0.707 [0.517, 0.965]* | 11.183 [7.712, 16.216]*** | 0.732 [0.536, 0.999]* | 2.072 [1.821, 2.358]*** | 4.258 [2.742, 6.612]*** | 1.665 [1.067, 2.600]* |
| Age | 1.078 [1.056,1.100]*** | 1.085 [1.070,1.100]*** | 1.138 [1.105,1.171]*** | 1.082 [1.064,1.101]*** | 1.154 [1.126,1.182]*** | 1.095 [1.042,1.152]*** | 1.122 [1.085,1.161]*** |
| Age$^2$ | 1.001 [1.000,1.001]** | 1.000 [1.000,1.001]** | 1.000 [0.999,1.001] | 1.000 [1.000,1.001] | 0.999 [0.998,1.000]** | 1.000 [0.999,1.002] | 0.999 [0.998,1.001] |
| Period (ref: 2010) | | | | | | | |
| 2011 | 1.032 [0.900,1.183] | 1.011 [0.884,1.157] | 0.686 [0.565,0.832]*** | 1.027 [0.818,1.290] | 0.872 [0.654,1.162] | 0.955 [0.716,1.274] | 0.916 [0.711,1.181] |
| 2012 | 0.785 [0.692,0.890]*** | 0.794 [0.716,0.881]*** | 0.425 [0.284,0.637]*** | 0.874 [0.792,0.964]** | 0.741 [0.513,1.069] | 0.666 [0.509,0.870]** | 0.626 [0.427,0.917]* |
| 2013 | 0.679 [0.581,0.795]*** | 0.720 [0.615,0.843]*** | 0.283 [0.210,0.380]*** | 0.688 [0.573,0.825]*** | 0.549 [0.439,0.687]*** | 0.628 [0.482,0.818]*** | 0.732 [0.475,1.127] |
| 2015 | 0.449 [0.390,0.516]*** | 0.430 [0.390,0.474]*** | 0.168 [0.130,0.217]*** | 0.445 [0.401,0.494]*** | 0.388 [0.307,0.491]*** | 0.288 [0.206,0.405]*** | 0.497 [0.354,0.697]*** |
| 2017 | 0.229 [0.183,0.287]*** | 0.230 [0.201,0.265]*** | 0.069 [0.048,0.101]*** | 0.248 [0.218,0.283]*** | 0.184 [0.162,0.208]*** | 0.150 [0.110,0.204]*** | 0.252 [0.162,0.393]*** |
| 2018 | 0.158 [0.125,0.201]*** | 0.193 [0.164,0.227]*** | 0.041 [0.027,0.063]*** | 0.187 [0.159,0.219]*** | 0.121 [0.103,0.143]*** | 0.099 [0.070,0.141]*** | 0.141 [0.092,0.214]*** |
| 2021 | 0.077 [0.050,0.120]*** | 0.106 [0.080,0.141]*** | 0.011 [0.007,0.017]*** | 0.125 [0.097,0.161]*** | 0.043 [0.034,0.055]*** | 0.031 [0.020,0.048]*** | 0.036 [0.019,0.066]*** |
| Sex (Female = 1) | 1.329 [1.253,1.410]*** | 1.371 [1.260,1.492]*** | 1.570 [1.448,1.702]*** | 1.337 [1.213,1.475]*** | 1.695 [1.532,1.874]*** | 1.299 [1.184,1.424]*** | 1.128 [0.995,1.279] |
| *Random effects* | | | | | | | |
| $\sigma^2$ Cohort | 0.094 [0.013, 0.654] | 0.042 [0.011, 0.169] | 0.001 [0.000, 0.057] | 0.015 [0.001, 0.164] | 0.000 [0.000, 0.000] | 0.060 [0.012, 0.309] | 0.008 [0.000, 7.396] |
| BIC | 52460.3 | 33932.1 | 12424.4 | 21900.3 | 11424.8 | 9820.3 | 6067.9 |
| AIC | 52380.8 | 33857.4 | 12360.8 | 21829.7 | 11357.2 | 9760.9 | 6014.2 |
| N | 50,721 | 29,715 | 21,006 | 18,698 | 13,479 | 12,451 | 6,093 |

*Note.* Age was centered at 45. Fixed-effect estimates are reported as odds ratios (ORs) with 95% confidence intervals. Models are unweighted, and standard errors are clustered at the cohort level. AIC = Akaike information criterion; BIC = Bayesian information criterion. *** $p < .001$, ** $p < .01$, * $p < .05$.

**Table S7 Unweighted Multilevel Logistic Regression of Digital Exclusion in CHARLS Stratified by Health Status**

| | No chronic disease | Single chronic disease | Multimorbidity | Cognitively not at risk | Cognitively at risk |
|---|---|---|---|---|---|
| Intercept | 10.218 [7.905, 13.209]*** | 8.538 [6.741, 10.815]*** | 11.535 [9.633, 13.813]*** | 8.122 [5.714, 11.543]*** | 49.420 [21.818, 111.941]*** |
| Age | 1.090 [0.963,1.234] | 1.112 [1.083,1.143]*** | 1.088 [1.068,1.109]*** | 1.014 [0.980,1.049] | 1.047 [0.868,1.263] |
| Age$^2$ | 1.001 [0.998,1.004] | 1.000 [0.999,1.002] | 1.000 [1.000,1.000] | 1.002 [1.001,1.003]*** | 1.001 [0.998,1.005] |
| Period (ref: 2011) | | | | | |
| 2013 | 0.579 [0.397,0.843]** | 0.705 [0.499,0.996]* | 0.565 [0.494,0.647]*** | 0.610 [0.541,0.687]*** | 0.527 [0.366,0.761]*** |
| 2015 | 0.323 [0.199,0.526]*** | 0.368 [0.277,0.489]*** | 0.389 [0.331,0.457]*** | 0.398 [0.328,0.482]*** | 0.319 [0.193,0.527]*** |
| 2018 | 0.136 [0.064,0.292]*** | 0.145 [0.120,0.175]*** | 0.136 [0.115,0.162]*** | 0.178 [0.134,0.238]*** | 0.070 [0.030,0.163]*** |
| 2020 | 0.024 [0.011,0.049]*** | 0.021 [0.016,0.028]*** | 0.023 [0.016,0.033]*** | 0.036 [0.027,0.049]*** | 0.012 [0.005,0.030]*** |
| Sex (Female = 1) | 1.488 [1.396,1.585]*** | 1.687 [1.563,1.822]*** | 1.770 [1.621,1.933]*** | 1.198 [1.098,1.307]*** | 1.638 [1.436,1.869]*** |
| *Random effects* | | | | | |
| $\sigma^2$ Cohort | 0.009 [0.000, 53.549] | 0.000 [0.000, 0.000] | 0.000 [0.000, 0.000] | 0.101 [0.030, 0.344] | 0.051 [0.000, 15.426] |
| BIC | 20488.4 | 11576.5 | 14280.0 | 28464.1 | 9000.0 |
| AIC | 20430.7 | 11520.8 | 14221.3 | 28396.9 | 8942.0 |
| N | 28,074 | 20,956 | 32,421 | 33,062 | 29,293 |

*Note.* Age was centered at 45. Fixed-effect estimates are reported as odds ratios (ORs) with 95% confidence intervals. Models are unweighted, and standard errors are clustered at the cohort level. Multimorbidity was defined as having two or more chronic conditions. Cognitive function was measured as a composite score (range: 1–36); a cutoff of 18 was used to classify respondents as at risk of cognitive impairment versus not at risk. AIC = Akaike information criterion, BIC = Bayesian information criterion. *** p < .001, ** p < .01, * p < .05.